\documentclass[
aip,
jcp,
showpacs,
amsmath,
amssymb,
floatfix,
longbibliography,
letterpaper,
lengthcheck,
superscriptaddress
]{revtex4-1}

\usepackage{graphicx}
\usepackage{hyperref}
\usepackage{braket}

\newcommand{\m}[1]{\boldsymbol{#1}}
\newcommand{\Z}{\mathcal{Z}}
\newcommand{\Tr}{\mathrm{Tr}}
\newcommand{\tr}{\mathrm{tr}}

\newcommand{\norm}[1]{\left|\left|#1\right|\right|}
\newcommand{\toop}[1][]{\mathcal{T}_{#1}}
\newcommand{\expval}[1]{\left\langle#1\right\rangle}

\begin{document}

\title{Many-body Green's function theory for electron-phonon interactions:\\
ground state properties of the Holstein dimer}

\author{Niko \surname{S{\"a}kkinen}}
%\email{niko.sakkinen@jyu.fi}
\affiliation{
Department of Physics, Nanoscience Center,
University of Jyv{\"a}skyl{\"a},
Survontie 9, 40014 Jyv{\"a}skyl{\"a}, Finland}

\author{Yang \surname{Peng}}
%\email[Electronic address:\;]{peng@fhi-berlin.mpg.de}
\affiliation{Dahlem Center for Complex Quantum Systems and Fachbereich Physik, Freie Universit{\"a}t Berlin, 14195 Berlin, Germany}
\affiliation{
Fritz-Haber-Institut der Max-Planck-Gesellschaft,
Faradayweg 4-6, D-14195 Berlin-Dahlem, Germany}

\author{Heiko \surname{Appel}}
%\email{appel@fhi-berlin.mpg.de}
\affiliation{
Fritz-Haber-Institut der Max-Planck-Gesellschaft,
Faradayweg 4-6, D-14195 Berlin-Dahlem, Germany}

\author{Robert \surname{van Leeuwen}}
%\email{robert.vanleeuwen@jyu.fi}
\affiliation{
Department of Physics, Nanoscience Center,
University of Jyv{\"a}skyl{\"a},
Survontie 9, 40014 Jyv{\"a}skyl{\"a}, Finland}

%%%%%%%%%%%%%%%%%%%%%%%%%%%%%%%%%%%%%%%%%%%%%%%%%%%%%%%%%%%%%%%%%%
%                            Abstract                            %
%%%%%%%%%%%%%%%%%%%%%%%%%%%%%%%%%%%%%%%%%%%%%%%%%%%%%%%%%%%%%%%%%%
%------------------------------------------------------------------------------%

\begin{abstract}
We study ground-state properties of a two-site, two-electron Holstein model
describing two molecules coupled indirectly via electron-phonon interaction
by using both exact diagonalization and self-consistent diagrammatic many-body
perturbation theory. The Hartree and self-consistent Born approximations used
in the present work are studied at different levels of self-consistency.
The governing equations are shown to exhibit multiple solutions when
the electron-phonon interaction is sufficiently strong whereas at smaller
interactions only a single solution is found. The additional solutions at larger
electron-phonon couplings correspond to symmetry-broken states with
inhomogeneous electron densities. A comparison to exact results indicates that
this symmetry breaking is strongly correlated with the formation of a bipolaron
state in which the two electrons prefer to reside on the same molecule.
The results further show that the Hartree and partially self-consistent Born
solutions obtained by enforcing symmetry do not compare well with exact
energetics, while the fully self-consistent Born approximation improves
the qualitative and quantitative agreement with exact results in the same
symmetric case. This together with a presented natural occupation number
analysis supports the conclusion that the fully self-consistent approximation
describes partially the bipolaron crossover. These results contribute
to better understanding how these approximations cope with the strong
localizing effect of the electron-phonon interaction.
\end{abstract}

% 31.15.xm quasi-particle methods
% 31.15.xp Perturbation Theory
% 71.10.Fd Lattice Fermion Models
% 71.38-k Polarons and electron-phonon interactions
% 71.38.Mx Bipolarons

\date{\today}
\maketitle

\noindent

%%%%%%%%%%%%%%%%%%%%%%%%%%%%%%%%%%%%%%%%%%%%%%%%%%%%%%%%%%%%%%%%%%
%                        Introduction                            %
%%%%%%%%%%%%%%%%%%%%%%%%%%%%%%%%%%%%%%%%%%%%%%%%%%%%%%%%%%%%%%%%%%

\section{Introduction}
\label{section:Introduction}
The electron-phonon interaction has proven to be an useful concept in many
systems for describing coupled motions of electron and nuclei. This model for
the interaction explains classic physical phenomena, such as conventional
superconductivity~\cite{schrieffer-book}, and plays an important role in many
active fields such as charge transport in molecular
junctions~\cite{galperin-2007,cuevas-book,dash-2010,dash-2011,wilner-2014}.
This field contains a wealth of interesting physics, negative differential
resistance~\cite{boese-2001,zazunov-2006}, hysteresis~\cite{alexandrov-2003},
image charge dynamics~\cite{myohanen-2012,thygesen-2014,perfetto-2013},
negative friction~\cite{kartsev-2014} and multistability~\cite{alexandrov-2003b,
galperin-2008,albrecht-2012,wilner-2013,khosravi-2012} to name a few, but is
at the same time a challenge for a theorist. The quantum transport problem calls
for a method which can cope with time-dependence in inhomogeneous open systems,
and with interactions between charge carriers and other constituents. In
previous work~\cite{myohanen-2008,myohanen-2009} we have developed
non-equilibrium many-body perturbation theory (MBPT) based on the Kadanoff-Baym
equations~\cite{stefanucci-book,dahlen-2007} to study time-dependent
quantum transport in systems with electron-electron interactions. The long
term goal is to study time-dependent transport phenomena in a single formalism
in the presence of both electron-phonon and electron-electron interactions.
As a first step towards this goal, we have included the electron-phonon
interactions in this formalism for finite systems in equilibrium and
non-equilibrium~\cite{saekkinen-2014b}. As a benefit of addressing finite
systems, we are given the possiblity to test the performance of many-body 
approximations by comparing to exact diagonalization~\cite{fehske-2007} results
which are not available for open systems. In the present work, we focus on
the electron-phonon interaction only to make this comparison as transparent
as possible, and to simplify the analysis of the used many-body approximations.
This also means that we can test the many-body approximations in a situation
where there are qualitatively different ground states depending on the strength
of the electron-phonon interaction. Such a situation is relevant in the case of 
quantum transport as the initial state affects the physics of quantum transport
and is related to phenomena such as bistability~\cite{alexandrov-2003b,
galperin-2008,albrecht-2012,wilner-2013}.

The finite system studied here is the homogeneous, two-site Holstein model
which is a minimalistic model representing interacting electrons and
phonons~\cite{holstein-2000}. The electrons are allowed to occupy localized
molecular orbitals and couple via their density to a primary molecular
vibrational mode. This electron-phonon interaction gives rise to a situation
in which the electron and nuclear motions are intertwined creating a bound
state known as a polaron~\cite{fehske-2007,devreese-2009}. The Holstein model
has been widely used to study extended systems such as molecular
crystals~\cite{hannewald1,hannewald2}. Also the ground-state properties
of the two-site system have been studied extensively with analytic and
numerically exact methods~\cite{schmidt-1987,feinberg-1990,ranninger-1992,
alexandrov-1994,marsiglio-1995,demello-1997,crljen-1998,rongsheng-2002,
qingbao-2005,paganelli-2008a, paganelli-2008b}, in various forms of
order-by-order perturbation theory~\cite{firsov-1997,chatterjee-2002}, and
in terms of a cumulant treatment~\cite{gunnarsson-1994}.
The present work extends these studies to diagrammatic self-consistent
many-body perturbation theory which we have previously applied to purely
electronic systems~\cite{dahlen-2005,dahlen-2007,saekkinen-2012}.
Although self-consistent many-body theory has been used ubiquitously
to investigate the ground state and spectral properties of the extended
model~\cite{marsiglio-1990,freericks-1994,capone-2003,goodvin-2006,bauer-2011}, 
to our knowledge small finite systems have not been investigated
to the same extent. This observation together with the facts that finite systems
form the interacting region of a typical quantum transport
calculation~\cite{cuevas-book}, and that the quality of a many-body
approximation can change considerably with the system
size~\cite{holm-1998,stan-2006,caruso-2013}, makes the investigation
of these approximations relevant also in the case of finite systems. The focus
of the present work is in the situation in which the electron-phonon coupling
leads to an effective attractive interaction between electrons, and for
a sufficiently strong interaction to a two-electron bound state known as
a bipolaron~\cite{fehske-2007,devreese-2009}. The competition between
de-localization and localization brought upon by the kinetic energy and
interaction is seen as qualitatively different ground states in the weak- and
strong-interaction limits. This work addresses the question how many-body
perturbation theory describes this change and the strong localizing effect of
the interaction when homogeneity is not enforced a priori.

The self-consistent diagrammatic many-body perturbation
theory~\cite{stefanucci-book} is used in the present work to obtain the dressed
electron and phonon propagators which can be used to calculate one-body
observables and interaction energies. These objects are obtained from coupled
Dyson equations self-consistently, meaning that the self-energy is a functional
of these propagators. This implies that perturbation theory is done to infinite
order in the interaction albeit that only certain classes of perturbative terms
are summed to infinite order. The self-energies are space-time non-local
potentials describing interactions between electrons and phonons, and are
subject to approximations. We are, in particular, interested in the so-called
conserving approximations~\cite{baym-1961,baym-1962,stefanucci-book}, which are
vital in quantum transport as they guarantee the conservation of energy, momentum
and particle number. The self-energy approximations considered are the Hartree
and partially or fully self-consistent Born approximations~\cite{galperin-2007}
which all are conserving in the sense that they conserve the particle number.
The partially self-consistent Born approximation is a standard
approximation in the quantum transport case, while its fully self-consistent
counterpart is not used as commonly~\cite{galperin-2007}, but has been studied
in the high-dimensional extended Holstein model e.g.~in combination with
dynamical mean-field theory~\cite{capone-2003,bauer-2011}. One aim of this work
is to bridge the gap between these two approximations by investigating
the effects of increased self-consistency realized by dressing the phonon
propagator. Another important open question is the behavior of these
approximations when we do not explicitly restrict ourselves to a homogeneous 
solution, but allow spontaneous symmetry-breaking, lack of which has been
attributed to the breakdown of a fully self-consistent Born approximation
in the extreme adiabatic limit~\cite{alexandrov-2001}.
As we will show, the different physical regimes of the system show up as
multiple solutions to the self-consistent equations of many-body perturbation
theory, and some of these solutions do break the reflection symmetry of
the two-site model. This is an example of the existence of multiple solutions
in many-body perturbation theory, a topic which has been addressed recently
in a more general context~\cite{tandetzky-2012}, but here these solutions have
a physical origin being related to a bipolaron formation.

The paper is organized as follows. First, we introduce many-body perturbation
theory and the framework in which it is used. This includes introducing
the self-energy approximations, and explaining how observables are calculated.
Then the two-site Holstein model is introduced, followed by the main section
on analytic and numerical results in which we analyze in detail the exact
bipolaronic ground state, and multiple solutions and symmetry breaking
in the approximations. The results are presented starting with an exact
diagonalization study, followed by results for different levels of perturbation
theory, and ending with a comparison to exact results. Lastly, we conclude with
an outlook and summary of the results. Supplementary material is presented
in the appendices.

%%%%%%%%%%%%%%%%%%%%%%%%%%%%%%%%%%%%%%%%%%%%%%%%%%%%%%%%%%%%%%%%%%
%                       System and Method                        %
%%%%%%%%%%%%%%%%%%%%%%%%%%%%%%%%%%%%%%%%%%%%%%%%%%%%%%%%%%%%%%%%%%

\section{Theory}
\label{Section:Theory}

\subsection{Hamiltonian}
\label{Section:Theory:Hamiltonian}

The present work is based on many-body perturbation theory. 
In this section we briefly introduce the physical context to which
we apply our approach. The physical system studied here is
a prototype example, namely a system of interacting fermions
and bosons or, as we will henceforth say, electrons and phonons.
The Hamiltonian operator of this system can, in general, be written as
\begin{align*}
   \hat{H}
   =&\sum_{i}\omega_{i}\hat{a}_{i}^{\dagger}\hat{a}_{i}
   +\sum_{ij}h_{ij}\hat{c}^{\dagger}_{i}\hat{c}_{j} \notag\\
   &+\sum_{ijk}\big(m_{jk}^{i}\hat{a}^{\dagger}_{i}
   +m_{kj}^{i{*}}{}\hat{a}_{i}\big)
   \hat{c}^{\dagger}_{j}\hat{c}_{k}\, ,
\end{align*}
where $\hat{c}_{i}/\hat{c}^{\dagger}_{i}$ are electron and
$\hat{a}_{i}/\hat{a}^{\dagger}_{i}$ phonon annihilation/creation
operators. These operators obey the usual canonical commutation
relations
\begin{align*}
   \{\hat{c}_{i},\hat{c}_{j}^{\dagger}\}
   =\delta_{ij}\, ,\\
   [\hat{a}_{i},\hat{a}^{\dagger}_{j}]
   =\delta_{ij}\, ,
\end{align*}
where $\delta_{ij}$ denotes the Kronecker delta.
The properties of the system are encoded in the elements of
the electron $h_{ij}$ and phonon $\omega_{i}\delta_{ij}$ one-body matrices,
and in the electron-phonon interaction tensor $m_{jk}^{i}$.

In order to be consistent with the standard representation
of many-body perturbation theory of interacting electrons
and phonons~\cite{mahan-book,bruus-book}, we further
introduce the self-adjoint phonon operators
\begin{align*}
   \hat{\phi}_{1,i}
   &\equiv\big(\hat{a}^{\dagger}_{i}+\hat{a}_{i}\big)/\sqrt{2}\, ,
   &\hat{\phi}_{2,i}
   &\equiv\imath\big(\hat{a}^{\dagger}_{i}-\hat{a}_{i}\big)/\sqrt{2}\, ,
\end{align*}
to which we associate a collective index
$I\equiv\{\varsigma_{i}\in\{1,2\},i\}$
so that we can write their commutation relation compactly as
\begin{align*}
   [\hat{\phi}_{I},\hat{\phi}_{J}]
   =\alpha_{IJ}\, ,
\end{align*}
where
\begin{align*}
   \alpha_{i\varsigma_{i},j\varsigma_{j}}
   &\equiv-\sigma_{2;\varsigma_{i}\varsigma_{j}}\delta_{ij}\, ,
\end{align*}
and
\begin{align*}
   \sigma_{2}
   =&\begin{pmatrix}
      0 & -\imath \\
      \imath & 0
   \end{pmatrix}
\end{align*}
denotes the second Pauli spin matrix. These
operators, proportional to the displacement and momentum field
operators, allow us to rewrite the Hamiltonian operator as
\begin{align}
\label{Eq:HamiltonianFieldRepresentation}
   \hat{H}
   =&\sum_{IJ}\Omega_{IJ}\hat{\phi}_{I}\hat{\phi}_{J}
   +\sum_{ij}h_{ij}\hat{c}^{\dagger}_{i}\hat{c}_{j}\notag\\
   &+\sum_{Ijk}M_{jk}^{I}\hat{\phi}_{I}\hat{c}^{\dagger}_{j}\hat{c}_{k}\, ,
\end{align}
where the new phonon and electron-phonon interaction elements are defined as
\begin{align*}
   \Omega_{i\varsigma_{i},j\varsigma_{j}}
   &\equiv\omega_{i}(\delta_{ij}\delta_{\varsigma_{i}\varsigma_{j}}
   +\alpha_{i\varsigma_{i},j\varsigma_{j}})/2\, ,\\
   M_{jk}^{i\varsigma_{i}}
   &\equiv\delta_{\varsigma_{i},1}\big(m_{jk}^{i}+m_{kj}^{i}{}^{*}\big)/\sqrt{2}
   \notag\\
   &-\imath\delta_{\varsigma_{i},2}\big(m_{jk}^{i}-m_{kj}^{i}{}^{*}\big)/\sqrt{2}
   \, .
\end{align*}

%------------------------------------------------------------------------------%

\subsection{Many-Body Perturbation Theory}
\label{Section:Theory:ManyBodyPerturbationTheory}
The electron and phonon propagators are the building blocks of
the many-body perturbation theory discussed here. They are functions which
describe the propagation of electrons and phonons in the interacting system
and can be used to evaluate expectation values of mostly one-body observables.
The central objects are the field expectation values
\begin{align}
\label{Eq:PhiDef}
   \phi_{I}(z)
   &\equiv\frac{1}{\Z}
   \Tr\bigg[\toop\Big\{e^{-\imath\int\!d\bar{z}\;\hat{H}(\bar{z})}
   \hat{\phi}_{I}(z)\Big\}\bigg] \, ,
\end{align}
electron propagators
\begin{align*}
   G_{ij}(z;z')
   &\equiv \frac{1}{\imath\Z}
   \Tr\bigg[\toop\Big\{e^{-\imath\int\!d\bar{z}\;H(\bar{z})}
   \hat{c}_{i}(z)\hat{c}^{\dagger}_{j}(z')\Big\}\bigg] \, ,
\end{align*}
and phonon propagators
\begin{align*}
   D_{IJ}(z;z')
   &\equiv \frac{1}{\imath\Z}
   \Tr\bigg[\toop\Big\{e^{-\imath\int\!d\bar{z}\;H(\bar{z})}
   \Delta\hat{\phi}_{I}(z)\Delta\hat{\phi}_{J}(z')\Big\}\bigg] \, ,
\end{align*}
where $\Delta\hat{\phi}_{I}\equiv\hat{\phi}_{I}-\phi_{I}$ is a fluctuation
operator, and all operators are given in the Schr\"{o}dinger picture
but have time-arguments $z,z',\bar{z}$ for book-keeping
reasons~\cite{stefanucci-book}. Moreover $\hat{H}(z)$ denotes
the Hamiltonian operator, $\Z\equiv\Tr[e^{-\imath\int\!dz\;\hat{H}(z)}]$
the grand-canonical partition function, $\Tr$ the trace over a complete
set of quantum states, and $\toop$ is the time-ordering operator on a Keldysh
time-contour~\cite{stefanucci-book}.

The perturbation expansion of these objects with respect to the electron-phonon
interaction can be constructed with help of Wick's theorem. The diagrammatic two
$G$-and $D$-line irreducible form of this perturbation theory allows us to write
these perturbation expansions in a closed form. In the present work,
we focus on the equilibrium properties which are accessible
in the imaginary-time, or Matsubara formalism. The contour times are defined
as $z\equiv-\imath\tau\, ,\tau\in[0,\beta]$, and propagators, which are
in general two-time functions, become functions of the relative time only
\begin{align*}
   a^{M}(\tau-\tau')\equiv a(z=-\imath\tau,z'=-\imath\tau')\, ,
\end{align*}
where $a$ is a two-time function, and the superscript $M$ refers
to a Matsubara component. The expectation values
of the field operators can then be written as
\begin{align}
\label{Eq:FieldExpectationValue}
   \phi^{M}_{i\varsigma_{i}}
   &\equiv\phi^{M}_{i\varsigma_{i}}(\tau)
   =\frac{\imath}{\omega_{i}}\sum_{jk}M_{jk}^{i\varsigma_{i}}G_{kj}^{M}(0^{-}),
\end{align}
where the superscript $0^{-}$ refers to taking the limit to zero from below.
The electron and phonon propagators satisfy non-linear Fredholm integral
equations of the second kind. These equations, known as Dyson equations, 
are given by
\begin{align*}
   \m{G}^{M}(\tau)
   =&\m{g}^{M}(\tau)
   +\big[\m{g}^{M}\star\m{\Sigma}^{M}\star\m{G}^{M}\big](\tau) \, ,\\
   \m{D}^{M}(\tau)
   =&\m{d}^{M}(\tau)
   +\big[\m{d}^{M}\star\m{\Pi}^{M}\star\m{D}^{M}\big](\tau) \, ,
\end{align*}
where boldfaced symbols denote matrices, with the usual definition
$(\m{a}\m{b})_{ij}\equiv \sum_{k}a_{ik}b_{kj}$ of a matrix product,
and the convolutions are defined as
\begin{align*}
   [\m{a}^{M}\star\m{b}^{M}](\tau)
   \equiv-\imath\int\limits_{0}^{\beta}\!d\tau'\;
   \m{a}^{M}(\tau-\tau')
   \m{b}^{M}(\tau')\, ,
\end{align*}
where $\m{a},\m{b}$ are matrix valued Matsubara functions.
The bare propagators, that is propagators of the non-interacting system,
which appear above are given by
\begin{align*}
   \m{g}^{M}(\tau)
   =&-\imath\theta(\tau)\big(1-f_{+}(\beta\m{h}^{M})\big)e^{-\m{h}^{M}\tau}\notag\\
   &+\imath\theta(-\tau)f_{+}(\beta \m{h}^{M})e^{-\m{h}^{M}\tau} \, ,\\
   \m{d}^{M}(\tau)
   =&-\imath\m{\alpha}\theta(\tau)
   \big(1+f_{-}(\beta\m{\alpha\omega})\big)e^{-\m{\alpha\omega}\tau}\notag\\
   &-\imath\m{\alpha}\theta(-\tau)f_{-}(\beta\m{\alpha\omega})e^{-\m{\alpha\omega}\tau}\, ,
\end{align*}
where $\m{h}^{M}\equiv\m{h}-\mu$ and $\mu$ is the chemical potential,
$(\m{\alpha\omega})_{i\varsigma_{i},j\varsigma_{j}}
\equiv \omega_{i}\alpha_{i\varsigma_{i},j\varsigma_{j}}$,
$\theta$ denotes the Heaviside function, and 
$f_{\pm}(z)\equiv\big(e^{z}\pm 1\big)^{-1}$ denote the Fermi-Dirac
and Bose-Einstein distribution functions, respectively.
The integral kernels $\Sigma\equiv\Sigma[G,D]$ and $\Pi\equiv\Pi[G,D]$
that appear in the Dyson equations are non-local one-body potentials
known as electron and phonon self-energies. The self-energies contain
information about interactions and are objects which need to be approximated.
The electron and phonon propagators of the interacting system
can be obtained once a self-energy approximation is chosen.

%------------------------------------------------------------------------------%

\subsection{Self-Energy Approximations}
\label{Section:Theory:Approximations}
\begin{figure}
   \centering
   \includegraphics[width=0.4\textwidth]{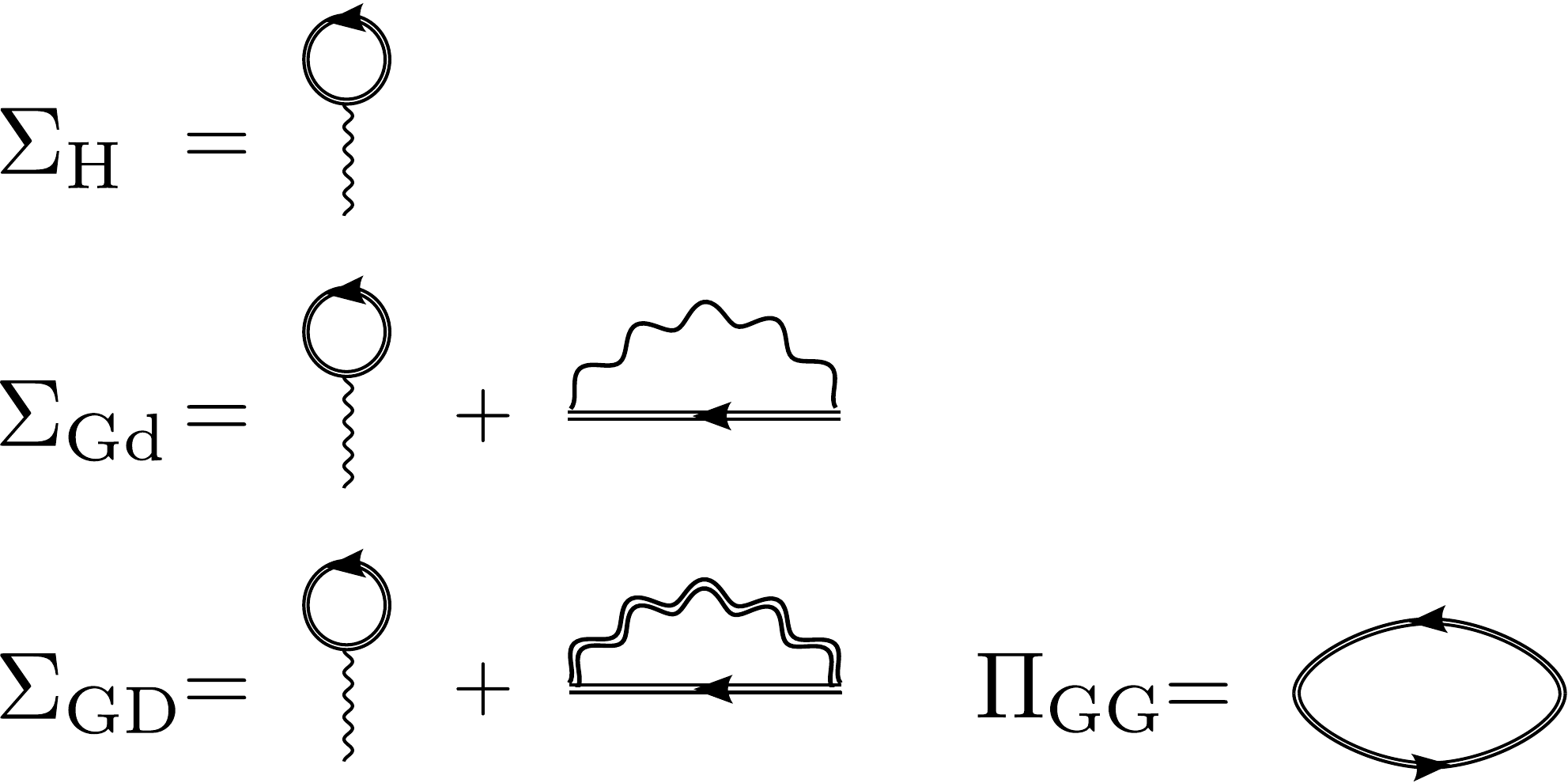}
   \caption{\label{Fig:Selfenergy}The Hartree (H),
   partially self-consistent (Gd), and fully self-consistent (GD) Born self-energies
   summarize the many-body approximations used in this work. A line with an arrow
   indicates a dressed electron propagator, while single and two-fold wiggly
   lines represent bare and dressed phonon propagators, respectively.}
\end{figure}

The self-energy approximations used in this work are shown diagrammatically
in Fig.~\ref{Fig:Selfenergy}. The top graph of this figure represents
the Hartree (H) approximation, while the middle graph is the partially
self-consistent Born (Gd) approximation~\cite{galperin-2007}. The bottom
graph represents the fully self-consistent Born (GD) approximation which is
also known as Migdal-Eliashberg (ME) approximation~\cite{migdal-1959,
eliashberg-1960a,eliashberg-1960b}.

The Hartree approximation is the simplest
approximation: the electron self-energy is approximated
with the Hartree diagram, which can be written as
\begin{align*}
   \Sigma_{\mathrm{H}}[G]^{M}_{ij}(\tau)
   &= \imath\delta(\tau)\sum_{K} M_{ij}^{K}\phi^{M}_{K}\, , 
\end{align*}
and the phonon self-energy is neglected. The Hartree approximation
is a time-local, static approximation which corresponds to a mean-field
description.

The second approximation is the partially self-consistent
Born approximation. This approximation also takes into
account the Fock (F) diagram
\begin{align*}
   \Sigma_{\mathrm{F}}[G,D]^{M}_{ij}(\tau)
   &=\imath\sum_{kl,PQ}
   M_{ik}^{P}M_{lj}^{Q}
   D^{M}_{PQ}(\tau) G^{M}_{kl}(\tau) \, , 
\end{align*}
which is a time-nonlocal memory term describing single-phonon
absorption/emission processes. The electron self-energy is given
in this approximation by
\begin{align*}
   \Sigma_{\mathrm{Gd}}[G]^{M}_{ij}(\tau)
   &\equiv\Sigma_{\mathrm{H}}[G]^{M}_{ij}(\tau)
   +\Sigma_{\mathrm{F}}[G,d]^{M}_{ij}(\tau)
   \, ,
\end{align*}
where $d$ is the bare phonon propagator which means that
the phonon self-energy is taken to be zero.

The third, and last, approximation is the fully self-consistent
Born approximation in which
the electron self-energy is given by
\begin{align*}
   \Sigma_{\mathrm{GD}}[G,D]^{M}_{ij}(\tau)
   &\equiv\Sigma_{\mathrm{H}}[G]^{M}_{ij}(\tau)
   +\Sigma_{\mathrm{F}}[G,D]^{M}_{ij}(\tau)
   \, ,
\end{align*}
and the phonon self-energy is approximated with
the bubble diagram 
\begin{align}
\label{Eq:bubble-selfenergy}
   \Pi_{\mathrm{GG}}[G]^{M}_{IJ}(\tau)
      &= -\imath\sum_{kl,pq}
      M_{kl}^{I}M_{pq}^{J}
      G^{M}_{kq}(\tau)G^{M}_{pl}(-\tau) \, ,
\end{align}
which describes simple phonon induced electron-hole excitation processes.
 
%------------------------------------------------------------------------------%

\subsection{Observables}
\label{Section:Theory:Observables}
The electron and phonon propagators allow us to evaluate one-body observables,
and additionally some more complicated observables such as the interaction 
energies. In the following, we introduce the observables considered
in this work, and explain how they can be evaluated from the knowledge
of the electron and phonon propagators.

The equilibrium electron one-body reduced density matrices are given by
\begin{align*}
   \gamma_{ij}
   &=\langle\hat{c}^{\dagger}_{j}\hat{c}_{i}\rangle\notag\\
   &=-\imath G^{M}_{ij}(0^{-})\, ,
\end{align*}
where the angular brackets denote the grand-canonical ensemble average
similar to Eq.~\eqref{Eq:PhiDef}.
The diagonal elements of this density matrix give
the electron density
\begin{align*}
   n_{i}
   &\equiv\gamma_{ii}\, ,
\end{align*}
while also off-diagonal elements are needed in order to evaluate electronic
natural occupation numbers and/or orbitals. In order to address energetics,
we consider the electron $E_{e}$, phonon $E_{p}$, and 
electron-phonon $E_{ep}$ interaction energies
which are evaluated according to
\begin{align*}
   E_{e}
   &=\sum_{ij}h_{ij}\gamma_{ji}
   \, ,\\
   E_{pC}
   &\equiv\sum_{IJ}\Omega_{IJ}\phi^{M}_{J}\phi^{M}_{I}
   \, ,\\
   E_{p}
   &=E_{pC}
   +\imath\sum_{IJ}\Omega_{IJ} D^{M}_{JI}(0^{-})
   \, ,\\
   E_{epC}
   &\equiv-\imath\sum_{Ijk}\phi_{I}^{M}M^{I}_{jk}G^{M}_{kj}(0^{-})
   \, ,\\
   E_{ep}
   &=E_{epC}-\sum_{ij}\int\limits_{0}^{\beta}\!d\tau\;
   \big(\Sigma^{M}_{ij}(-\tau)-\Sigma^{M}_{H;ij}(-\tau)\big)G^{M}_{ji}(\tau)
   \, ,
\end{align*}
where the additional subscript $C$ refers to the classical
(mean-field) piece of this energy. The total energy is then evaluated
by summing all contributions according to
\begin{align*}
   E\equiv E_{e}+E_{p}+E_{ep} \, .
\end{align*}
This is in agreement with the energy functional
\begin{align*}
   E_{\mathrm{GM}}
   &= \imath\sum_{i}\partial_{\tau}G^{M}_{ii}(\tau)\big|_{\tau=0^{-}} \notag\\
   &-\frac{1}{\imath}\sum_{IJ}\Omega_{IJ}\big(D_{JI}^{M}(0^{-})
   -\imath\phi_{J}^{M}\phi_{I}^{M}\big) \, ,
\end{align*}
which is known, in the purely electronic case, as the Galitski-Migdal
functional. The derivation of this functional is given
in the appendix~\ref{Appendix:EnergyFunctional}.

%%%%%%%%%%%%%%%%%%%%%%%%%%%%%%%%%%%%%%%%%%%%%%%%%%%%%%%%%%%%%%%%%%
%                         Model                                  %
%%%%%%%%%%%%%%%%%%%%%%%%%%%%%%%%%%%%%%%%%%%%%%%%%%%%%%%%%%%%%%%%%%

\section{Model}
\label{section:Model}
Our model system is a two-site Holstein model~\cite{
schmidt-1987,feinberg-1990,ranninger-1992,alexandrov-1994,
marsiglio-1995,demello-1997,crljen-1998,rongsheng-2002,qingbao-2005,
paganelli-2008a,paganelli-2008b,firsov-1997,chatterjee-2002} which can
be viewed as a minimal representation of a system in which electrons
move between two molecules, so that they are coupled to the vibrational
modes of these molecules. The Hamiltonian operator for a single primary
vibrational mode of two identical molecules, henceforth referred to as
sites 1 and 2, and minimal, localized basis for electrons is given by
\begin{align*}
   \hat{H}
   &=\omega_{0}\sum_{i=1}^{2}\hat{a}_i^\dagger\hat{a}_i
   -g\sum_{i=1}^{2}(\hat{a}^{\dagger}_i+\hat{a}_i)\hat{n}_i
   + \hat{H}_{e}  \, ,\\
   \hat{H}_{e}
   &=-T\sum_\sigma\left(\hat{c}_{1\sigma}^\dagger\hat{c}_{2\sigma}
   +\hat{c}_{2\sigma}^{\dagger} \hat{c}_{1\sigma}  \right)
\end{align*}
where $\hat{a}_{i}$ is the phonon annihilation operator at site $i$,
$\hat{c}_{i\sigma}$ is the electronic operator that annihilates
an electron of spin $\sigma$ at site $i$, and
$\hat{n}_i=\sum_{\sigma}\hat{c}^{\dagger}_{i\sigma}\hat{c}_{i\sigma}$
is the electron density operator at site $i$.
The parameters $\omega_{0}$, $T$ and $g$ characterize the bare vibrational
frequency, inter-site hopping and local electron-phonon interaction strength,
respectively. The displacement and momentum operators, defined in this
model as $\hat{u}_{i}\equiv\hat{\phi}_{1,i}$ and
$\hat{p}_{i}\equiv\hat{\phi}_{2,i}$, allow us to rewrite the Hamiltonian
operator as
\begin{align*}
   \hat{H}
   =&\frac{\omega_{0}}{2}\sum_{i=1}^{2}
   \big(\hat{p}_{i}^{2}+\hat{u}_{i}^{2}-1\big)
   -\sqrt{2}g\sum_{i=1}^{2}\hat{u}_{i}\hat{n}_{i}
   +H_{e} \, ,
\end{align*}
which is equivalent to the Hamiltonian of
Eq.~\eqref{Eq:HamiltonianFieldRepresentation} with the matrix elements
$\Omega_{i\varsigma_{i},j\varsigma_{j}}
=\omega_{0}\delta_{ij}
(\delta_{\varsigma_{i}\varsigma_{j}}
-\sigma_{2;\varsigma_{i}\varsigma_{j}})/2$
and
$M_{j\sigma,k\sigma'}^{i\varsigma_{i}}
=-\sqrt{2}g\delta_{\varsigma_{i},1}
\delta_{\sigma\sigma'}\delta_{ij}\delta_{jk}$. The properties of
this model depend on two parameters, the adiabatic ratio
\begin{align*}
   \gamma\equiv\frac{\omega_{0}}{T} \, ,
\end{align*}
and the effective interaction
\begin{align}
\label{eq.effective_interaction}
   \lambda
   &\equiv\frac{2g^2}{T\omega_{0}} \, .
\end{align}
The adiabatic ratio $\gamma$ describes the relative energy scale of
electrons and nuclei, while the effective interaction $\lambda$
is a measure of the coupling between the motions of these two constituents.
This interaction is equal to the ratio of the Lang-Firsov~\cite{langfirsov-1962}
bipolaron energy to the energy of two free electrons, and turns out to
be an useful quantity for analyzing the many-body approximations.

%%%%%%%%%%%%%%%%%%%%%%%%%%%%%%%%%%%%%%%%%%%%%%%%%%%%%%%%%%%%%%%%%%
%                         Results                                %
%%%%%%%%%%%%%%%%%%%%%%%%%%%%%%%%%%%%%%%%%%%%%%%%%%%%%%%%%%%%%%%%%%

\section{Results}
\label{section:Results}

%------------------------------------------------------------------------------%

\subsection{Exact Properties}
\label{Section:Results:ExactProperties}
The main feature of the Holstein model is that it undergoes a cross-over,
as a function of the interaction strength, from nearly free electrons
to self-trapped quasi-particles known as polarons~\cite{fehske-2007,
devreese-2009}. Our focus is in the two-electron case in which
a phonon-mediated attractive interaction leads to the formation of
a polaron pair referred to as a bipolaron. The purpose of this section is to 
clarify what is this correlated state and how it behaves as a function of
the adiabatic ratio and effective interaction.

Our discussion starts with a limiting case obtained by rewriting
the Hamiltonian operator as
\begin{align*}
   \hat{H}
   =&\omega_{0}\sum_{i=1}^{2}
   \bigg(\hat{a}_{i}^{\dagger}-\frac{g}{\omega_{0}}\hat{n}_{i}\bigg)
   \bigg(\hat{a}_{i}-\frac{g}{\omega_{0}}\hat{n}_{i}\bigg)\notag\\
   &-\frac{g^{2}}{\omega_{0}}\sum_{i=1}^{2}\sum_{\sigma}
   \hat{c}^{\dagger}_{i\sigma}\hat{c}_{i\sigma}
   -\frac{2g^{2}}{\omega_{0}}\sum_{i=1}^{2}
   \hat{n}_{i\uparrow}\hat{n}_{i\downarrow}
   +\hat{H}_{e} 
\end{align*}
and considering the situation $2g^2/\omega_0 \gg T$, namely $\lambda \gg 1$.
The electron Hamiltonian $\hat{H}_{e}$ has the energy scale $T$ and is thus
a candidate for a perturbation expansion, assuming that a finite converge
radius exists. The unperturbed Hamiltonian obtained by setting $T$ to zero is 
diagonal in the electron and phonon site representations. The first term in this 
operator describes two shifted harmonic oscillators whose shift depends on
the electron population and interaction strength. The second term is
an electronic one-body term, and the last term represents an attractive
interaction between electrons. The eigenstates can be labeled with eigenvalues
$n_{i}$ of the site density operators $\hat{n}_{i}$ since they commute with
the Hamiltonian. Then the degenerate two-electron ground state of
the unperturbed system obtained by choosing $n_1=2,~n_2=0$ or $n_2=2,~n_1=0$ 
consists of a localized electron pair accompanied by a lattice distortion
described by a shifted oscillator. This state is to be understood in
this case as a bipolaron, that is a bound state of two electrons and a lattice
displacement.

At finite hopping $T$, the electronic Hamiltonian $\hat{H}_e$ removes
the degeneracy, and the competition between the de-localizing effect of
the kinetic energy and the localizing effect of the interaction determines
whether or not the ground state can be characterized with this quasi-particle.
As we will soon see, in this case, the notion of a bipolaron is not perfect
as there is no sharp distinction between a nearly-free electron and
a bipolaronic state. In the following, we consider this regime and give
a more precise definition of a bipolaron. This discussion requires some
preliminaries starting with the transformation
\begin{align*}
   \hat{a}
   &\equiv(\hat{a}_1 - \hat{a}_2) / \sqrt{2} \, , \\
   \hat{A}
   &\equiv(\hat{a}_1 + \hat{a}_2 ) / \sqrt{2} \, ,
\end{align*}
where operators $\hat{a}$ and $\hat{A}$ fulfill the fundamental commutation
relations for bosons. The original Hamiltonian can then be separated into two 
parts, $\hat{H} = \hat{H}_{tot} + \hat{H}_{rel}$, describing the total and
relative motion of the dimer respectively. In particular,
\begin{align}
\label{Eq:TotalHamiltonian}
   \hat{H}_{tot}
   &\equiv\omega_0\hat{A}^\dagger\hat{A}-g\hat{N}\hat{U} \, , \\
\label{Eq:TotalDisplacement}
	\hat{U}
   &\equiv(\hat{A}+\hat{A}^\dagger)/\sqrt{2} \, ,
\end{align}
where $\hat{U}$ is the operator for the total displacement. This Hamiltonian
simply corresponds to a shifted harmonic oscillator which can be solved exactly. 
The relative part of the Hamiltonian is given by
\begin{align}
\label{Eq:RelativeHamiltonian}
   \hat{H}_{rel}
   &\equiv\omega_{0}\hat{a}^\dagger\hat{a}
   -g(\hat{n}_1-\hat{n}_2)\hat{u}+\hat{H}_{e}\\
\label{Eq:RelativeDisplacement}
   \hat{u}
   &\equiv(\hat{a}+\hat{a}^\dagger)/\sqrt{2} \, ,
\end{align}
where $\hat{u}$ is the operator for the relative displacement. The eigenvalue
equation for the full Hamiltonian $\hat{H}$ can be written as
\begin{align*}
   \hat{H} \ket{\Phi^{N}_{(n,j)}}
   &=E^N_{(n,j)}\ket{\Phi^{N}_{(n,j)}} \, ,\\
\end{align*}
where we denote the $N$-electron eigenstates as $\ket{\Phi^{N}_{(n,j)}}$ and 
their energies as $E^N_{(n,j)}$. The double labeling $(n,j)$ reflects the fact
that, because of the partitioning of $\hat{H}$ into two parts, the full
eigenstates can be written as the products
\begin{align*}
   \ket{\Phi^{N}_{(n,j)}}
   &=\ket{n,N}\ket{\Psi^{N}_{j}}\, ,  \\
   \hat{H}_{tot}\ket{n,N}
   &=\left(\omega_{0}n
   -\frac{g^2N^2}{2\omega_{0}}\right) \ket{n,N}\, , \\ 
   \hat{H}_{rel}\ket{\Psi^{N}_{j}}
   &=\varepsilon_n^N\ket{\Psi^{N}_{j}}\, ,
\end{align*}
where $\ket{n,N}$ and $\ket{\Psi^N_{j}}$ are the eigenstates of
$\hat{H}_{tot}$ and $\hat{H}_{rel}$ respectively.
The Hamiltonian commutes with the total spin $\hat{S}^2$ and z component
$\hat{S}_z$, $[\hat{S}^2,\hat{H}]=0$ and $[\hat{S}_z,\hat{H}]=0$, therefore
we can classify states according to their spin angular momentum.
The ground state for two electrons ($N=2$) is in the subspace with
$S(S+1)=0$ and $S_{z}=0$. The relative Hamiltonian is also invariant under
the operation of the parity operator, which swaps indices $1$ and $2$ and
changes $\hat{u}$ to $-\hat{u}$. Hence the eigenstates of $\hat{H}_{rel}$ have 
either even or odd parity. In particular, the ground state is even under parity 
operation. Thus by symmetry, we have $\langle\hat{u}\rangle$=0,
$\langle\hat{n}_1\rangle=\langle \hat{n}_2 \rangle=1$, 
which are expectation values taken with respect to the symmetric ground state.
\begin{figure*}
	\centering
   \includegraphics[width=\textwidth]{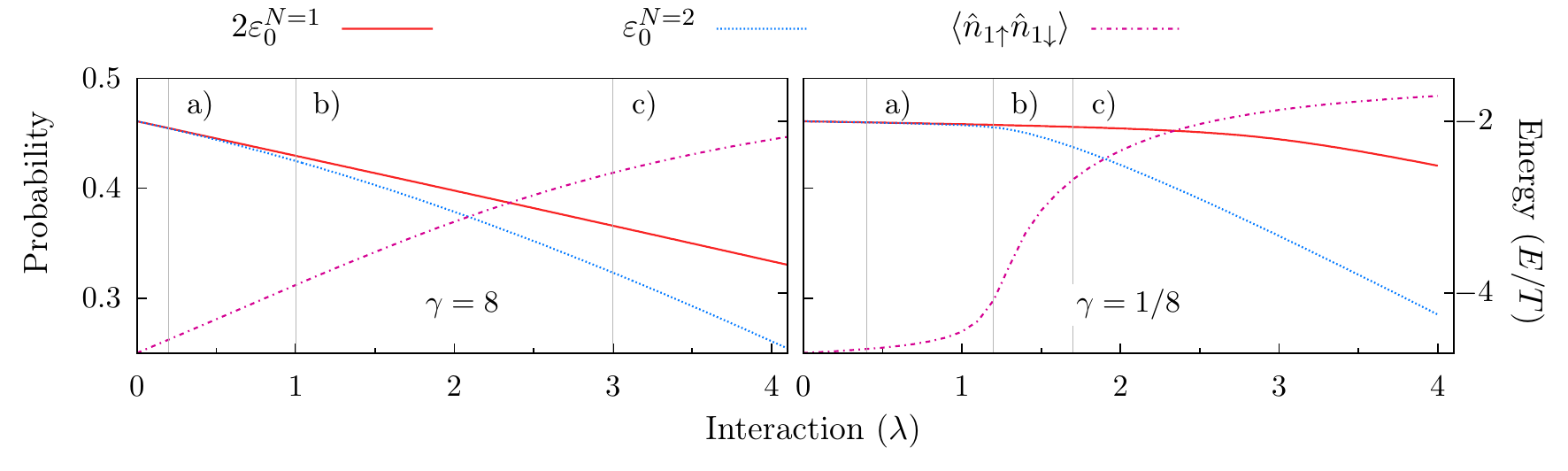}
   \includegraphics[width=\textwidth]{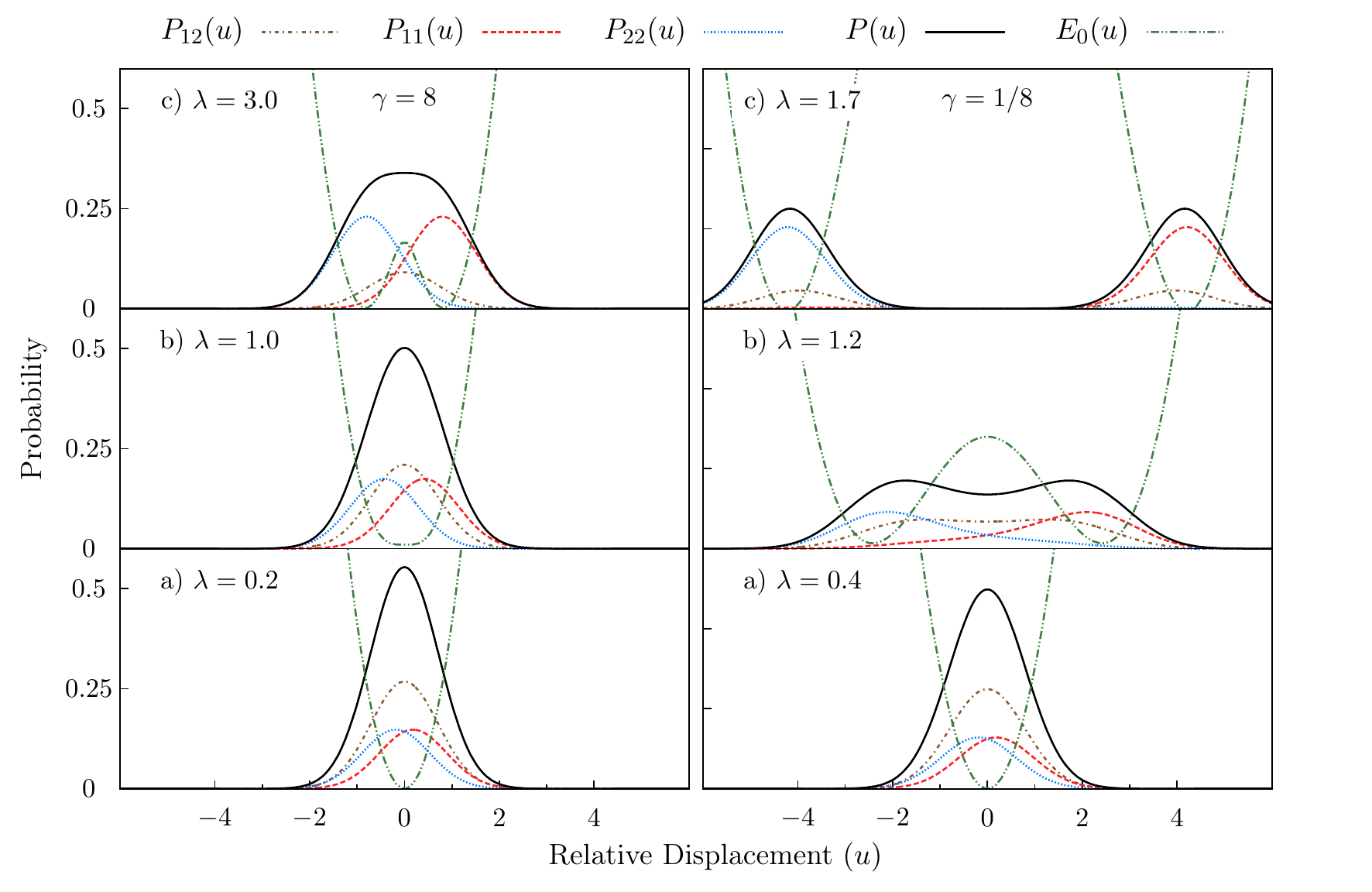}
   \caption{\label{fig.probability} The top panel displays the two-electron
   $\varepsilon^{N=2}_0$, two times one-electron $2\varepsilon^{N=1}_0$
   ground-state energies, and double occupancy
   $\langle \hat{n}_{1\uparrow}\hat{n}_{1\downarrow}\rangle$ as
   a function of electron-phonon interaction strength $\lambda$.
   The energies are represented on the right (Energy) and double-occupancies on
   the left (Probability) axis, respectively. The bottom panel shows the
   joint probabilities $P_{ij}(u)$ to find electrons at the sites $i,j$
   and nuclei at the relative displacement $u$, and the probability $P(u)$ to
   find the nuclear displacement at $u$ for three different electron-phonon
   interaction strengths. The lowest BO surfaces $E_0$ for
   different cases are shown in the same panels. They are plotted
   according to Eq.\eqref{eq.BO_surface}
   with only a constant shift, in order
   to let their minima touch the bottom of the figures. Left:
   Small hopping regime, $\gamma=8$ in the anti-adiabatic limit.
   Right: Large hopping regime,  $\gamma=1/8$ in the adiabatic limit.
   (color online)}
\end{figure*}
The two-electron singlet Hilbert space is then spanned by
\begin{align}
   \ket{1}_e
   &=\frac{1}{\sqrt{2}}\left(
   \hat{c}_{1\uparrow}^\dagger\hat{c}_{2\downarrow}^\dagger
   -\hat{c}_{1\downarrow}^\dagger\hat{c}_{2\uparrow}^\dagger \right)
   \ket{0}_{e}\, ,\nonumber \\
   \ket{2}_e
   &=\frac{1}{\sqrt{2}}\left(
   \hat{c}_{1\uparrow}^\dagger\hat{c}_{1\downarrow}^\dagger
   +\hat{c}_{2\uparrow}^\dagger\hat{c}_{2\downarrow}^\dagger \right)
   \ket{0}_{e}\, , \label{eq.basis}\\
   \ket{3}_e
   &=\frac{1}{\sqrt{2}}\left(
   \hat{c}_{1\uparrow}^\dagger\hat{c}_{1\downarrow}^\dagger
   - \hat{c}_{2\uparrow}^\dagger\hat{c}_{2\downarrow}^\dagger \right)
   \ket{0}_{e}\, ,\nonumber
\end{align}
where $\ket{0}_{e}$ denotes the electronic vacuum. The relative Hamiltonian can
be written in this subspace as the operator-valued matrix 
\begin{align}
\label{eq:block}
   \m{\hat{H}}_{rel}^{N=2}=
   \begin{pmatrix}
	   \omega_0\hat{a}^{\dagger}\hat{a} & -2T & 0\\
      -2T & \omega_0\hat{a}^{\dagger}\hat{a} & -2g\hat{u} \\
	   0 & -2g\hat{u} & \omega_0\hat{a}^{\dagger}\hat{a}
   \end{pmatrix} \, .
\end{align}
This Hamiltonian can be diagonalized numerically by expressing the phonon
operators as matrices in a basis set of Fock number states
$\left\{\ket{n}_{ph}\right\}$, which in practical calculations is truncated up
to a maximum value $N_{ph}$, i.e. $n\leq N_{ph}$. In our exact diagonalization 
approach $N_{ph}=100$ turns out to be sufficient for the convergence of the relevant
quantities. Alternatively, using the fact that
\begin{align}
   \hat{a}^\dagger \hat{a}
   &=\frac{\hat{p}^2}{2} +\frac{\hat{u}^2}{2} -\frac{1}{2}\, ,
\label{eq.real_space}
\end{align}
and by choosing a finite difference representation
for momentum $\hat{p}$ and the displacement $\hat{u}$ operators,
the diagonalization can also be performed in real space. 
Either of the two diagonalization approaches yield accurate ground-state  energies
as long as the truncated basis for nuclear motion is large enough.
However, each approach has its own advantages in calculating certain properties.

We are now set to investigate the finite $T$ properties of this model
and illustrate the bipolaronic nature of the system.
In Fig.~\ref{fig.probability}, we show results for two sets of parameters, 
corresponding to the small and large electron hopping regimes: $\gamma=8$
in anti-adiabatic regime on the left hand side and $\gamma=1/8$ in adiabatic
regime on the right hand side. The top panels show the two-electron
ground-state energies $\varepsilon_{0}^{N=2}$, that is the lowest eigenvalues of
$\m{\hat{H}}_{rel}^{N=2}$,  as a function of the effective electron-phonon
interaction $\lambda$ for both adiabatic ratios $\gamma$. These ground-state 
energies are compared with two times the one-electron ground-state energies
$2\varepsilon_{0}^{N=1}$, which are obtained by  diagonalizing
the relative Hamiltonian represented in the one-electron subspace with
$S_z=1/2$ (same as $S_z=-1/2$). The figure shows that when the electron
phonon interaction strength is small, the energies are very close to one
another, and when the electron-phonon interaction strength is increased,
the two-electron ground-state becomes much lower in energy and decreases faster
as a function of the interaction. This suggests a physical picture in which
for weak interactions the two electrons, or polarons, are almost independent.
A strong electron-phonon interaction, on the other hand, gives rise to
a large effective attraction between electrons leading to a strongly bound
electron, or polaron pair, which is seen as increased binding energy
$2\varepsilon_{0}^{N=1}-\varepsilon_{0}^{N=2}$. This picture can be elucidated by
investigating the double occupancy which is a correlation function of
the occupations of spin up and down electron at the same site. Without loss of
generality, we focus on site 1 for the double occupancy
\begin{align*}
   \langle\hat{n}_{1\uparrow}\hat{n}_{1\downarrow}\rangle
   &= \bra{\Psi_0^{2}}\hat{n}_{1\uparrow}
   \hat{n}_{1\downarrow}
   \ket{\Psi_0^{2}},
\end{align*}
where $\hat{n}_{1\alpha}=\hat{c}_{1\alpha}^\dagger \hat{c}_{1\alpha}$, 
and $\ket{\Psi_0^{2}}$ is the two-electron ground state. The double-occupancy
is shown in the top panels of Fig.~\ref{fig.probability}
as a function of the electron phonon interaction strength $\lambda$ for both
adiabatic ratios $\gamma$. The figure shows that this quantity is equal
to one-quarter for a non-interacting system stating due to symmetry that
it is equally probable to find electrons in the same of different sites.
As the interaction is increased, the double-occupancy approaches one-half 
irrespectively of the adiabatic ratio. This result implies that an electron-pair
is formed as a function of the interaction. The effect of the adiabatic ratio
is that the crossover from the non-interacting system to a system in which
an electron pair is formed happens more abruptly in the adiabatic $\gamma=1/8$
case.

The increase of the binding energy indicates a crossover to a strongly bound
two-electron state when the interaction is strong enough. This goes hand in hand
with an increase of the double-occupancy which says that the probability of
finding two electrons at the same site is large compared to the probability
of finding them at different sites. This picture can be made more quantitative
by introducing the joint probability of finding electrons at the sites $i,j$
and nuclei at the relative coordinate $u$. The joint probability is defined as
\begin{align}
	P_{ii}(u)
   &\equiv\lvert\langle i\uparrow,i\downarrow;u\vert\Psi_{0}^{2}\rangle\rvert^{2}
   \, ,\notag\\
	P_{ij}(u)
   &\equiv\sum_{\substack{\sigma\sigma'\\\sigma\neq\sigma'}}
   \lvert\langle i\sigma,j\sigma';u\vert\Psi_{0}^{2}\rangle\rvert^{2}
   \, ,i\neq j \, , 
   \label{eq.joint_probability}
\end{align}
where
$\lvert i\sigma,j\sigma';u\rangle
\equiv\hat{c}^{\dagger}_{i\sigma}\hat{c}^{\dagger}_{j\sigma'}
\lvert 0\rangle_{e}\lvert u\rangle$
with $\lvert u\rangle$ being an eigenstate of $\hat{u}$.
The lower panels a), b) and c) of Fig.~\ref{fig.probability} display these
probabilities. The probability of finding electrons on the same site is in
the non-interacting limit equal to the probability of finding them at
different sites. As the electron-phonon 
interaction is increased, the maximum of the joint probability of finding
electrons on the site one (two) increases and the position of the maximum moves
towards larger (smaller) values of the relative displacement. On the other hand,
the maximum of the joint probability of finding electrons on neighboring sites
decreases and the position of this maximum remains fixed at the origin as the interaction
is increased. Thus for any non-zero interaction it is more likely to find
the electrons occupying the same site and accompanied by a non-zero displacement.
This however does not mean that the ground state is characterized well by
the notion of an electron pair and an associated displacement. This leads us to
the concept of a working definition of a bipolaron. We regard for the present
work a bipolaron being a good representative of the ground state when
the maxima of the joint probabilities of electrons appearing at the same site,
$\max_{u} P_{ii}(u), i=1,2$, cf.~Eq.~\eqref{eq.joint_probability},
is at least two times larger than the joint probability to finding the electrons
at different sites $\max_{u} P_{12}(u)$. We note that this definition
is not unique, and depending on ones viewpoint other choices can be made,
but it does denote a situation in which an electron pair accompanied by
a displacement is a dominant feature of the ground state.
This working definition allows us to summarize the results of
Fig.~\ref{fig.probability} by saying that panels a) and c) represent in both
adiabatic and anti-adiabatic regimes two nearly-free electrons and a bipolaron,
respectively. The middle panel b), on the other hand, corresponds to a ground
state which is not characterized well by a single notion of either two
nearly-free electrons, two polarons, or a bipolaron.

The lower panels of Fig.~\ref{fig.probability} also display
the probability
\begin{align}
   P(u)
	\equiv \sum_{i} P_{ii}(u)+P_{12}(u)
   \label{eq.nuclei_probability}
\end{align}
of finding the nuclei at relative coordinate $u$. The figure shows that prior
to the bipolaron formation, this probability distribution has a single maximum
located at the origin. As the interaction is increased, the ground state becomes 
more and more bipolaron-like  and the nuclear distribution $P(u)$ starts to
split symmetrically, as it should due to the symmetry of the Hamiltonian.
In the adiabatic case, it already forms two well-separated peaks when
$\lambda=1.7$ in panel c), indicating that it is more likely to find the nuclei
at $u$ far from zero, either positive or negative. In the anti-adiabatic case,
this splitting happens much slower as a function of $\lambda$ and the peaks are
still not well-separated at $\lambda=3$. However, we already see the trend of
the splitting behavior which is the same as in the adiabatic case, the only
difference is that a quantitatively much larger $\lambda$ is needed to split
the nuclear distribution. Hence, the formation of the bipolaron coincides with
the splitting of the nuclear distribution in the adiabatic limit, while in
the anti-adiabatic case, the bipolaron formation does not necessarily imply
the splitting $P(u)$, and there is a smaller nuclear displacement associated
with the bipolaron compared to the displacement in the adiabatic case.

As a last topic on exact properties, we discuss the adiabatic limit which
leads us naturally to the Born-Oppenheimer (BO) approximation.
This perspective leads us to scrutinize the splitting of the nuclear probability
$P(u)$ using the adiabatic potential energy surfaces.
We use Eq.~\eqref{eq.real_space}, and set $\hat{p}\equiv 0$ in
the BO approximation. Then we diagonalize the 3 by 3 block
in Eq.~\eqref{eq:block}, where $u$ is now taking the role of a parameter.
The eigenvalues $E_0(u)$, $E_1(u)$, and $E_2(u)$ correspond to the ground-state,
first, and second excited-state BO
surfaces~(Fig.~\ref{fig.probability}) respectively
\begin{subequations}
\begin{align}
   E_0(u)
   &=\omega_0 \left(\frac{u^2}{2}-\frac{2}{\gamma}
   \sqrt{1+\frac{\lambda\gamma}{2}u^2} \right) \, ,
\label{eq.BO_surface}
\end{align}
\begin{align}
   E_1(u)
   &=\frac{u^2}{2}\omega_0 \, ,
\label{eq.E1_surface}
\end{align}
\begin{align}
   E_2(u)&= 
   \omega_0 \left(\frac{u^2}{2}+\frac{2}{\gamma}
   \sqrt{1+\frac{\lambda\gamma}{2}u^2} \right) .
\label{eq.E2_surface}
\end{align}
\end{subequations}
In the BO or adiabatic approximation the nuclei are moving on the ground-state
surface $E_0(u)$, which behaves as an effective potential. The shape of this
potential determines therefore to a large extent the shape of $P(u)$. From
the figure, we clearly see that the formation of a double minimum in $E_0(u)$
directly attributes to the splitting of $P(u)$. In particular, in the adiabatic
regime $\gamma=1/8$, this correspondence is a very good approximation because
the kinetic energy of the electrons is much larger than the phonon frequency,
and hence the ground-state BO surface is enough to capture the motion of the
nuclei. In Eq.~\eqref{eq.BO_surface}, we notice that the second term in
the parenthesis, which is just the energy difference between the ground and
the first excited BO surfaces (Eq.~\eqref{eq.E1_surface}), is small when
$\gamma$ is large (anti-adiabatic). Hence in the anti-adiabatic regime, 
the coupling between the lowest two surfaces is large, especially near
the origin $u=0$. This is why in Fig. \ref{fig.probability} b) on the left,
the correspondence between the potential surface and the shape of
the probability distribution is not convincible. The reason is that the first
excited surface also influences the nuclear motion and considering only
the ground BO surface is not enough. 

From Eq.~\eqref{eq.BO_surface} it can also be worked out that the condition for
the formation of a double minimum, at
$u_{1,2}=\pm\sqrt{2(\lambda^2-1)/\lambda\gamma}$, is for $\lambda > 1$ simply given
by taking the derivative of $E_0(u)$ with respect to $u$. We can write down
the ground-electronic state $\ket{\chi_0^{BO}(u)}$ in the BO approximation
in terms of the basis defined in Eq.~\eqref{eq.basis} 
\begin{align*}
    \ket{\chi_{0}^{BO}(u)}
   =\frac{\ket{1}_e
   +\sqrt{1+\frac{\lambda\gamma}{2}u^2}\ket{2}_e
   +\sqrt{\frac{\lambda\gamma}{2}}u\ket{3}_e}
   {(2+\lambda\gamma u^2)^{1/2}} .
\end{align*}
Using this state, we can compute the double occupancy at site $1$ in the BO
approximation
\begin{align*}
    \langle \hat{n}_{1\uparrow} \hat{n}_{1\downarrow}\rangle 
    &= \int_{-\infty}^{\infty}\mathrm{d}u P_{BO}(u)
    \bra{\chi_0^{BO}(u)}\hat{n}_{1\uparrow}
    \hat{n}_{1\downarrow}\ket{\chi_0^{BO}(u)} \\
    &=\int_{-\infty}^{\infty}\mathrm{d}u P_{BO}(u)
    \frac{1+\lambda\gamma u^2 +2\sqrt{\frac{\lambda\gamma}{2}}
    \sqrt{1+\frac{\lambda\gamma}{2}u^2}u}{2(2+\lambda\gamma u^2)} \\
    &=\int_{-\infty}^{\infty}\mathrm{d}u P_{BO}(u)
    \frac{1+\lambda\gamma u^2}{2(2+\lambda\gamma u^2)},
\end{align*}
where $P_{BO}(u)$ is the nuclear distribution $P(u)$ in the BO picture. We have
used the fact the BO state is factorizable and $P_{BO}(u)$ is an even function
of $u$. Next, we consider the case when $\gamma \rightarrow 0$. If $\lambda<1$,
$E_{0}$ reaches its minimum at 0. We perform the harmonic approximation by
calculating 
\begin{align*}
   \frac{1}{\omega_0}\frac{d^{2}E_0}{du^{2}}\bigg|_{0}
   =1-\lambda, 
\end{align*}
which means $P_{BO}(u)$ is peaked at the origin with width
$\sim (1-\lambda)^{-1/4}$, independent of $\gamma$. For $\lambda>1$ the barrier
between the double minimum at $u_{1,2}$ becomes infinite. Hence, we can consider
each well separately and perform the harmonic approximation around $u_{1,2}$.
By taking the second derivative of $E_{0}(u)$ with respect to $u$ and evaluating
it at the two minima $u_{1,2}$, we obtain 
\begin{align*}
\frac{1}{\omega_{0}}\frac{d^{2}E_{0}}{du^{2}}\bigg|_{u_{1,2}}
=\frac{\lambda^2-1}{\lambda^2}.
\end{align*}
Thus for $\lambda>1$, $P_{BO}$ can be approximated by two Gaussian wave packets
around the positions $u_{1,2}$ with width
$\sim \left[\lambda^2/(\lambda^2-1)\right]^{1/4}$, independent of $\gamma$.
Moreover, in the limit $\gamma \rightarrow 0$, the second term in the integrand
changes very slowly near $u_{max}$, either $0$ or $u_{1,2}$, at which $P_{BO}$
is peaked. This is due to the fact that the rate of change is proportional to
$\sqrt{\gamma}\rightarrow 0$. Hence we can replace the term $\lambda\gamma u^2$
by $\lambda\gamma u_{max}^2$, for either $\lambda<1$ or $\lambda>1$, namely
\begin{align*}
    \langle \hat{n}_{1\uparrow} \hat{n}_{1\downarrow}\rangle
    &\simeq \frac{1+\lambda\gamma u_{max}^2}{2(2+\lambda\gamma u_{max}^2)} 
    \int_{-\infty}^{\infty}\mathrm{d}u P_{BO}(u) \\
    &=
    \begin{cases}
        1/4 & \lambda<1\\
        \frac{2\lambda^{2}-1}{4\lambda^{2}} & \lambda>1.
    \end{cases}
\end{align*}
This demonstrates the sharp change, a kink at $\lambda=1$, of the double
occupancy for an increasing electron-phonon interaction in the adiabatic case, 
and its value is always between $1/4$ and $1/2$. As a final remark, we emphasize
that the appearance of the double-well structure for $\lambda>1$ in the BO
picture is closely related to the issue of symmetry-breaking discussed in
the following sections.

%------------------------------------------------------------------------------%

\subsection{Mean-Field}
\label{Section:Results:MeanFieldSolution}
Let us begin scrutinizing the Holstein dimer from the perspective of
our many-body approximations by considering the simplest self-consistent method,
the Hartree approximation. This self-energy approximation gives rise
to a mean-field approximation in which the electrons feel the nuclear displacement
instantaneously. The displacement expectation value
\begin{align*}
   u_{i}
   &=\sqrt{2}gn_{i}/\omega_{0} \, ,
\end{align*}
allows us to write the Hartree self-energy as
\begin{align}
\label{eq:hartree}
   \Sigma_{\mathrm{H};ij}^{M}(\tau)
   &=-\imath\delta_{ij}\delta(\tau)T\lambda n_{i} \, ,
\end{align}
where $n_{i}$ is the spin-summed site density. This term is time-local
which means that it can be accounted for with an effective one-body Hamiltonian
given by
\begin{align}
\label{Eq:HamiltonianHartree}
   \m{h}_{\mathrm{H}}^{M}\equiv
   \begin{pmatrix}
      -T\lambda n_{1}-\mu & -T \\
      -T & -T\lambda n_{2}-\mu
   \end{pmatrix} \, ,
\end{align}
where $\mu$ is the chemical potential which is to be chosen in the following.
We refer to the nonlinear eigenvalue equations associated with this matrix
as the Hartree equations. The two eigenvalues and eigenvectors of
these equations are given by
\begin{align*}
   \epsilon_{\mathrm{H},\pm}
   &=-\frac{T\lambda N}{2}\pm T\sqrt{1+\lambda^{2}(n_{1}-N/2)^{2}}-\mu \, ,\\
   \psi_{\mathrm{H};1,\pm}
   &=\frac{-1}
   {\sqrt{1+(\lambda n_{1}+\mu/T+\epsilon_{\mathrm{H},\pm}(n_{1})/T)^{2}}} \, ,\\
   \psi_{\mathrm{H};2,\pm}
   &=\frac{\lambda n_{1}+\mu/T+\epsilon_{\mathrm{H},\pm}(n_{1})/T}
   {\sqrt{1+(\lambda n_{1}+\mu/T+\epsilon_{\mathrm{H},\pm}(n_{1})/T)^{2}}}\, ,
\end{align*}
where $N$ is the electron number and $\lambda$ the dimensionless
interaction of Eq.~\eqref{eq.effective_interaction}. The eigendecomposition
can be used to form a self-consistency relation for the density $n_{1}$
at the first site. This relation can be written for two electrons ($N=2$)
by fixing the chemical potential at
\begin{align}
\label{Eq:ChemicalPotential}
   \mu&=-T\lambda \, ,
\end{align}
and taking the zero-temperature limit ($\beta \gg 2T)$ so that it is given by
\begin{align*}
   n_{1}
   &=2|\psi_{\mathrm{H};1,-}|^{2} \notag\\
   &=\frac{2}{1+\big(\lambda(n_{1}-1)
   -\sqrt{1+\lambda^{2}(n_{1}-1)^{2}}\big)^{2}}
   \, .
\end{align*}
By reordering terms this can be written as a cubic equation for
the spin-summed site density at site $1$, which has
the three solutions
\begin{align*}
   n_{s}
   &\equiv1 \, ,\\
   n_{a\pm}
   &\equiv1\pm\sqrt{1-1/\lambda^{2}} \, ,
\end{align*}
where subscript $s$ and $a$ stand for symmetric and asymmetric, respectively.
The symmetric solution ($n_{s}$) is always real valued and hence
an acceptable candidate for the lowest energy solution. The asymmetric
solutions ($n_{a\pm}$) are however complex when $\lambda<1$ and therefore
acceptable solutions only when they become real valued, that is
for $\lambda\geq 1$. This means that at the critical interaction
$\lambda_{C}^{\mathrm{H}}\equiv1$ a single physically acceptable solution 
splits into three acceptable solutions.
Our next task is then to check which one is the lowest energy solution. 
This can be achieved by using the spin-summed reduced density matrices
\begin{align*}
   \gamma_{s}
   &\equiv
   \begin{pmatrix}
      1 & 1 \\
      1 & 1
   \end{pmatrix}
   \, ,\\
   \gamma_{a\pm}
   &\equiv
   \begin{pmatrix}
      n_{a\pm} & 1/\lambda \\
      1/\lambda & 2-n_{a\pm}
   \end{pmatrix}
\end{align*}
where the asymmetric density matrix is only defined for $\lambda \geq 1$.
These density matrices together with the displacement expectation value,
and momentum expectation value ($p_{i}=0$) can be used to evaluate
the electron $E_{e}$, phonon $E_{p}$, and electron-phonon $E_{ep}$
interaction energies. The energy components of the symmetric solution are given
\begin{align*}
   E_{e,s}/T&=-2\, , \\ 
   E_{p,s}/T&=\lambda\, ,\\ 
   E_{ep,s}/T&=-2\lambda \, ,
\end{align*}
and those of the asymmetric solution by
\begin{align*}
   E_{e,a\pm}/T&=-2/\lambda\, , \\ 
   E_{p,a\pm}/T&=2\lambda-1/\lambda\, ,\\ 
   E_{ep,a\pm}/T&=-4\lambda+2/\lambda \, .
\end{align*}
These different energy components allow us to evaluate
the total energies 
\begin{align}
\label{Eq:HartreeTotalEnergy}
   E_{s}/T
   &=-\lambda-2\, , \\
   E_{a\pm}/T
   &=-2\lambda-1/\lambda \, ,
\end{align}
which indicate that the asymmetric solution is the lowest energy solution
which by definition is the ground state.

As discussed in Sec.~\ref{Section:Results:ExactProperties},
for the ground state of the exact solution, the double-occupancy on a given site
approaches one-half for large interactions ($\lambda$) representing a bipolaron
state. On the other hand, in the mean-field approach, we have found that 
above the critical interaction $\lambda_{C}^{\mathrm{H}}$, the ground state
becomes degenerate with two symmetry-broken states. In these states
the electrons prefer to populate the same site, and as the displacement
is linearly proportional to the electron density at that site, 
this electron pair is accompanied by a lattice deformation.
This supports the conclusion that the mean-field approach mimics
the formation of a bipolaron by symmetry-breaking and localization.
The formation appears continuously, although in contrast to the exact
case not smoothly, and very rapidly, in the manner of a bifurcation. 

%------------------------------------------------------------------------------%

\subsection{Beyond Mean-Field, Asymptotic Solution}
\label{Section:Results:BeyondMeanFieldAsymptoticSolution}
Symmetry-breaking is a well-known feature of mean-field
theories~\cite{helgaker-book,reimann-2002,ivanov-2002}. This leads us to
question what is needed to remedy this situation. Let us explore this thought
by scrutinizing the partially self-consistent Born approximation. Although we
are not in a position to be able to handle this approximation purely
analytically, we can investigate its asymptotic behavior in the adiabatic
and anti-adiabatic limits.

The self-energy is a sum of the Hartree and Fock self-energies.
The former is given in Eq.~\eqref{eq:hartree} and the latter can be written as
\begin{align*}
   \Sigma_{\mathrm{F}}[G]^{M}_{ij}(\tau)
   &= \imath\omega_{0}T\lambda \delta_{ij}d(\tau)G^{M}_{ii}(\tau) \, ,
\end{align*}
where we introduced
\begin{align*}
   2\imath d(\tau)
   &=\theta(\tau)
   \big[f_{-}(\beta\omega_{0})e^{\omega_{0}\tau}
   +\big(1+f_{-}(\beta\omega_{0})\big)e^{-\omega_{0}\tau}\big]\notag\\
   &+\theta(-\tau)
   \big[f_{-}(\beta\omega_{0})e^{-\omega_{0}\tau}
   +\big(1+f_{-}(\beta\omega_{0})\big)e^{\omega_{0}\tau}\big] \, .
\end{align*}
as the diagonal elements of the displacement-displacement component of
the bare phonon propagator.

Let us start with the adiabatic limit in which $\omega_{0}\ll T$ independent
of $\lambda$. In this limit, the electron propagator decays much faster
than the phonon propagator allowing us to approximate
the latter with a constant value. The $\omega_{0}\rightarrow 0$ limit
gives the constant
\begin{align*}
   \imath\omega_{0}d(\tau)
   &\rightarrow \frac{1}{\beta} \, ,
\end{align*}
which can be used to write the self-energy
in the imaginary-frequency representation as
\begin{align*}
   \Sigma_{\mathrm{Gd}}[G]^{M}_{ij}(p_{n})
   &= \Sigma_{\mathrm{H}}[G]^{M}_{ij}
   +\delta_{ij}\frac{T\lambda}{\beta}G^{M}_{ii}(p_{n}) \, ,
\end{align*}
where $p_{n}=\imath\pi(2n+1)/\beta$ is a fermionic imaginary frequency.
The electron propagator then satisfies the Dyson equation
\begin{align*}
   \m{G}^{M}(p_{n})
   &=\m{g}^{M}(p_{n})
   +\m{g}^{M}(p_{n})
   \m{\Sigma}_{\mathrm{Gd}}[G]^{M}(p_{n})
   \m{G}^{M}(p_{n})
\end{align*}
which is diagonal in the frequencies, and one can see that
in the zero-temperature limit $\beta\gg 1$ the self-energy
approaches the Hartree self-energy. This analysis suggests that
the partially self-consistent Born approximation reduces
to the mean-field approximation in the adiabatic limit, and is thus 
similarly expected to break the symmetry at $\lambda=1$.

The anti-adiabatic limit $\omega_{0}\gg T$, independent of $\lambda$,
can be treated similarly. Now the phonon propagator decays rapidly
compared to the electron propagator and is non-negligible only
for $|\tau|\rightarrow 0$ or $\beta$. This allows us to approximate
the phonon propagator with its value at the limit $\omega_{0}\rightarrow\infty$
which amounts to
\begin{align*}
   \frac{\omega_{0}e^{\mp\omega_{0}\tau}}{1-e^{-\beta\omega_{0}}}
   &\rightarrow \delta_{\pm}(\tau)\, ,
\end{align*}
where we introduced a delta function defined as 
$\delta_{\pm}(\tau)\equiv \delta(\tau^{\mp})$ on the intervals
$[0,\beta]$ and $[-\beta,0]$, respectively, and zero otherwise.
The phonon propagator can therefore be written as
\begin{align*}
   2\imath\omega_{0}d(\tau)
   &=\delta_{+}(\beta-\tau)+\delta_{+}(\tau)\notag\\
   &+\delta_{-}(-\beta-\tau)+\delta_{-}(\tau) \, ,
\end{align*}
where $\tau\in[-\beta,\beta]$.
This approximation enters the self-energy which
in turn appears as an integral kernel in the convolution
\begin{align*}
   \big[\m{\Sigma}_{\mathrm{F}}^{M}\star\m{G}^{M}\big]_{ij}(\tau)
   &=-\imath\sum_{k}\int\limits_{0}^{\beta}\,d\tau'\;
   \Sigma_{\mathrm{F}}[G]_{ik}^{M}(\tau-\tau')G^{M}_{kj}(\tau') \notag\\
   &=-\imath\frac{T\lambda}{2}
   \big(G^{M}_{ii}(0^{+})+G^{M}_{ii}(0^{-})\big)G^{M}_{ij}(\tau),
\end{align*}
where $\tau$ is restricted on the open interval $(0,\beta)$
since the values of this convolution are different at the end points.
The convolution appears as an integral kernel
in the Dyson equation which allows us to disregard the discontinuity
at the end points. Then the identity
$G_{ij}^{M}(0^{+})-G_{ij}^{M}(0^{-})=-\imath\delta_{ij}$ allows us to
conclude that
\begin{align*}
   \Sigma_{\mathrm{Gd}}[G]^{M}_{ij}(\tau)
   =& \Sigma_{\mathrm{H}}[G]^{M}_{ij} \notag\\
   &+\delta_{ij}\delta(\tau)T\lambda G^{M}_{ii}(0^{-})
   -\imath\delta_{ij}\delta(\tau)\frac{T\lambda}{2} \notag\\
   =& \frac{\Sigma_{\mathrm{H}}[G]^{M}_{ij}}{2}
   +\imath\delta_{ij}\delta(\tau)\frac{\mu}{2} \, ,
\end{align*}
which shows that the self-energy reduces to the Hartree self-energy
plus a constant contributing to the chemical potential.
We conclude that in the anti-adiabatic limit, the partially self-consistent
Born approximation becomes again identical to the mean-field
but now with a renormalized interaction $\lambda/2$. This new interaction means
that also the bifurcation point shifts to $\lambda=2$. We therefore
expect that partial self-consistency does not prevent symmetry-breaking and
that, unlike in the Hartree approximation, the bifurcation
point is not independent of the adiabatic ratio.

%------------------------------------------------------------------------------%

\subsection{Beyond Mean-Field, Phonon Vacuum Instability}
\label{Section:Results:PhononVacuumInstability}
The discussion on the properties of the nuclear system has so far remained
at a mean-field level. In the following we investigate how the phonon propagator
is affected by the interaction. The phonon vacuum instability is a known
issue when the propagator is dressed with a bare polarization bubble, that is
one evaluated with non-interacting electron propagators~\cite{alexandrov-1994,
alexandrov-book}. Next, we revise this peculiarity in the context of the Holstein
dimer, and investigate what happens when we use a symmetry-broken instead of
a symmetric electron propagator.

The phonon propagator satisfies in the imaginary-frequency representation
the equation
\begin{align*}
   \m{D}^{M}(\nu_{n})
   &=\m{d}^{M}(\nu_{n})
   +\m{d}^{M}(\nu_{n})
   \m{\Pi}[G]^{M}(\nu_{n})
   \m{G}^{M}(\nu_{n}),
\end{align*}
where $\nu_{n}=\imath 2\pi n/\beta$ is a bosonic imaginary-frequency.
Observables are then accessible by either using analytic continuation to real
frequencies, or by transforming back to imaginary-time and taking the limit
$0^{-}$ to evaluate time-local observables. The Lehmann representation,
or a partial fraction decomposition, of the phonon propagator convenient for
either of the two choices can be accomplished by calculating the roots of
the characteristic function
\begin{align*}
   \det\big(\nu_{n}^{2}\m{1}
   -\omega_{0}^{2}\m{1}
   -\omega_{0}\m{\mathsf{\Pi}}[G]^{M}(\nu_{n})\big)\, ,
\end{align*}
where $\m{1}$ denotes the identity matrix,
and $\mathsf{\Pi}[G]^{M}_{ij}(\nu_{n})\equiv\Pi[G]^{M}_{1i,1j}(\nu_{n})$
the displacement-displacement component of the phonon self-energy.
The roots of this equation are usually understood via analytic continuation
as the renormalized phonon frequencies.

The imaginary-frequency representation of the self-energy given
in Eq.~\eqref{Eq:bubble-selfenergy} can be written as
\begin{align*}
   \mathsf{\Pi}[G]_{ij}^{M}(\nu_{n})
   &= \frac{2T\omega_{0}\lambda}{\beta}\sum_{q=-\infty}^{\infty}
   G_{ij}^{M}(p_{q})G_{ji}^{M}(p_{q}-\nu_{n})
   \, .
\end{align*}
The electron propagator is approximated with the mean-field propagator 
\begin{align*}
   \m{G}_{\mathrm{H}}^{M}(p_{n})
   &= \big(p_{n}\m{1}-\m{h}_{\mathrm{H}}^{M}\big)^{-1} \, ,
\end{align*}
where $\m{h}_{\mathrm{H}}^{M}\equiv\m{h}_{\mathrm{H}}^{M}(\lambda/\lambda_{C})$
is defined in Eq.~\eqref{Eq:HamiltonianHartree}, and the critical interaction
$\lambda_{C}\geq 1$ has been introduced to be able to represent also
the partially self-consistent propagator, which we have shown to reduce to
the mean-field one but with a renormalized interaction. This approximation
leads for a sufficiently low temperature $\beta\gg 2T$ to the self-energy
\begin{align*}
   \mathsf{\Pi}[G_{\mathrm{H}}]_{ij}^{M}(\nu_{n})
   &= 
   \frac{2T^{2}\omega_{0}\lambda\bar{\lambda}^{-1}}
   {\nu_{n}^{2}-4\bar{\lambda}^{2}T^{2}}
   \begin{pmatrix}
      1 & -1 \\
      -1 & 1
   \end{pmatrix}
   \, ,
\end{align*}
where the effective interaction
\begin{align*}
   \bar{\lambda}
   \equiv\bar{\lambda}(\lambda)
   =\theta(\lambda-\lambda_{C})\lambda/\lambda_{C}
   +\theta(\lambda_{C}-\lambda)
\end{align*}
is introduced to allow us to incorporate both the symmetric
($\lambda<\lambda_{C}$) and asymmetric ($\lambda>\lambda_{C}$) cases
in a single equation. The characteristic function is then evaluated
with this self-energy and set to zero which leads to the two equations
\begin{align*}
   0&=x^{2}-\omega_{0}^{2} \, ,\\
   0&=x^{2}-\omega_{0}^{2}
   -\frac{4T^{2}\omega_{0}^{2}\lambda\bar{\lambda}^{-1}}
   {x^{2}-4\bar{\lambda}^{2}T^{2}} \, ,
\end{align*}
where $x$ denotes the sought solution. The roots of the first equation are
just the bare phonon frequencies $\omega_{0}$ which reflects the fact that
the total displacement $U$ couples only to the electron number and therefore its
frequency remains invariant. The second equation can be re-written as a fourth
order polynomial equation whose roots $\omega_{\pm},-\omega_{\pm}$ satisfy
\begin{align*}
   \frac{\omega_{\pm}^{2}}{T^{2}}
   &=2\big(\gamma^{2}/4+\bar{\lambda}^{2}\big)\Bigg(1 
   \pm\sqrt{1+
   \frac{\gamma^{2}\bar{\lambda}^{-1}(\lambda-\bar{\lambda}^{3})}
   {\bar{\lambda}\big(\gamma^{2}/4+\bar{\lambda}^{2}\big)^{2}}}\Bigg)\, .
\end{align*}
These roots represent the renormalized phonon frequencies related to
the relative displacement $u$. The important observation here is that these
frequencies should be real valued, but in fact $\omega_{-}$ becomes imaginary
when $\lambda-\bar{\lambda}^{3}>0$ which indicates the phonon vacuum
instability. The critical value of interaction at which this happens is
\begin{align*}
   \lambda>1
\end{align*}
for a symmetric $\bar{\lambda}=1$ electron propagator, and
\begin{align*}
   \lambda<\lambda_{C}^{3/2}
\end{align*}
for an asymmetric $\bar{\lambda}=\lambda/\lambda_{C}$ electron propagator.
This shows that there is a finite region $\lambda\in(1,\lambda_{C}^{3/2})$
in which imaginary phonon frequencies appear.

The imaginary frequencies have a direct impact on observables which are
obtained, for example, using the displacement-displacement component
$\mathsf{D}^{M}_{ij}(\nu_{n})\equiv D_{1i,1j}^{M}(\nu_{n})$
of the phonon propagator. This component can be written as
\begin{align*}
   \m{\mathsf{D}}^{M}(\nu_{n})
   &=\omega_{0}\Big(
   \nu_{n}^{2}\m{1}-\omega_{0}^{2}\m{1}
   -\omega_{0}\m{\mathsf{\Pi}}[G_{\mathrm{H}}]^{M}(\nu_{n})\Big)^{-1} \, ,
\end{align*}
which shows that it is diagonalized by the same transformation as the self-energy.
The two diagonal components of this propagator representing the total and
relative displacement modes are given by
\begin{align*}
   \mathsf{D}_{U}^{M}(\nu_{n})
   =&d_{0}(\nu_{n}) \, ,\\
   \mathsf{D}_{u}^{M}(\nu_{n})
   =&\frac{\omega_{0}} {\omega_{+}^{2}-\omega_{-}^{2}}
   \sum_{j\in\{\pm\}}j\omega_{j}^{-1}
   \big(\omega_{j}^{2}-4\bar{\lambda}^{2}T^{2}\big)
   \mathsf{d}_{j}(\nu_{n})\, ,
\end{align*}
where the displacement-displacement component of the bare
phonon propagator is given by
\begin{align*}
   \mathsf{d}_{j}(\nu_{n})
   \equiv \frac{\omega_{j}}{\nu_{n}^{2}-\omega_{j}^{2}}
\end{align*}
for a mode labeled with $j$. This representation shows that
when the frequency $\omega_{-}$ approaches zero, the relative component diverges
due to the $1/\omega_{-}$ behavior, and when it becomes imaginary it also diverges
for $\omega_{-}=\imath2\pi n/\beta$ due to the divergence of the bare propagator. 

These results suggest in the mean-field case ($\lambda_{C}=1$) that enforcing
symmetry leads to instability at $\lambda=1$, while allowing asymmetry avoids it,
retaining a physically acceptable situation. This observation complements
the adiabatic picture in which a double well is formed for $\lambda>1$, leaving
the mean-field approximation the option to break the symmetry by falling into
one of the two minima, or to retain it and to lead to phonon mode softening.
Additionally, we have seen that adding partial self-consistency leads
to the renormalized interaction $\lambda/2$ in
the anti-adiabatic limit which implies that the bifurcation point of
the electron propagator is $\lambda_{C}=2$ in our analysis. This means that
dressing the phonon propagator leads to the phonon vacuum instability
for $\lambda\in(1,2\sqrt{2})$. As a consequence of this instability
the phonon propagator and therefore observables obtained from it diverge.

%------------------------------------------------------------------------------%

\subsection{Beyond Mean-Field, Full Numerical Solution}
\label{Section:Results:MultipleSolutionsBeyondMeanField}
As we have shown, the Hartree approximation gives rise to multiple solutions
and to a symmetry-broken lowest energy solution. We have also shown
that going one step beyond this approximation neither prevents
the appearance of multiple solutions nor symmetry-breaking in the extreme
adiabatic limits. Here, we address the question what happens for finite
adiabatic ratios, and moreover we investigate if dressing the phonon propagator
leads to qualitatively different results, as we have speculated in the previous
section.

\begin{figure}
   \centering
   \includegraphics[width=0.275\textwidth]{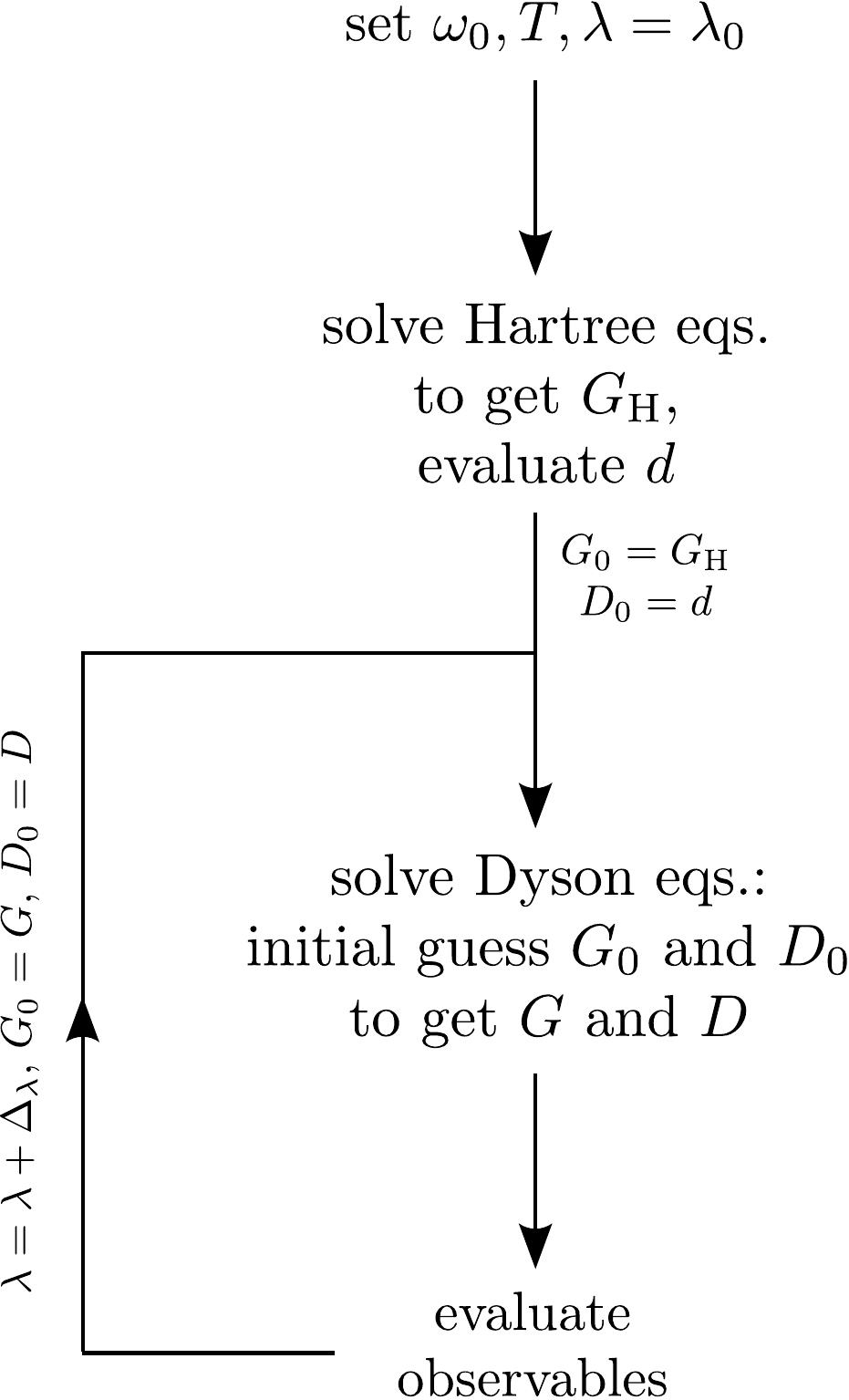}
   \caption{\label{Fig:Algorithm}
   Overview of the iterative process used to solve for the electron ($G$) and
   phonon ($D$) propagators. The symmetric solution is obtained by choosing
   $\lambda_{0}=0,\Delta_{\lambda}=0.01$ and the asymmetric solutions are
   obtained by choosing $\lambda_{0}=4,\Delta_{\lambda}=-0.01$. (color online)}   
\end{figure}

In the following, we compare ground-state results obtained
by exact diagonalization (ED) and approximate results obtained using
the Hartree (H), and partially (Gd) and fully (GD) self-consistent
Born approximations. The approximate results are calculated by choosing
the chemical potential of Eq.~\eqref{Eq:ChemicalPotential},
and the inverse temperature $\beta/\omega_{0}^{-1}=10^{3}$ which represents
the zero-temperature limit. Our objective is to show whether symmetric
and asymmetric solutions (co-)exist in all approximations for parameters
which span the adiabatic ($\gamma<1$) and anti-adiabatic ($\gamma>1$)
weak ($\lambda<1$) and strong ($\lambda>1$) coupling regimes. The adiabatic
ratio is varied in the present work by fixing $\omega_{0}$ to unity, and
changing only values of the hopping $T$ amplitude. The approximate results
are obtained by numerically solving the coupled electron and phonon Dyson
equations by using either a symmetric or an asymmetric initial
guess according to the iterative procedure summarized graphically 
in Fig.~\ref{Fig:Algorithm}. This procedure enforces the anti-hermiticity of
the propagators in the spatial indices and hence it does not allow for
unphysical solutions which would violate this property. We terminate
the self-consistent cycle once the residual norm of both propagators is below
a tolerance of $10^{-8}$ after which we say that we have converged
to a solution. The numerical details are discussed
in Appendix~\ref{Appendix:NumericalDetails}.
 
\begin{figure*}
   \centering
   \includegraphics[width=\textwidth]{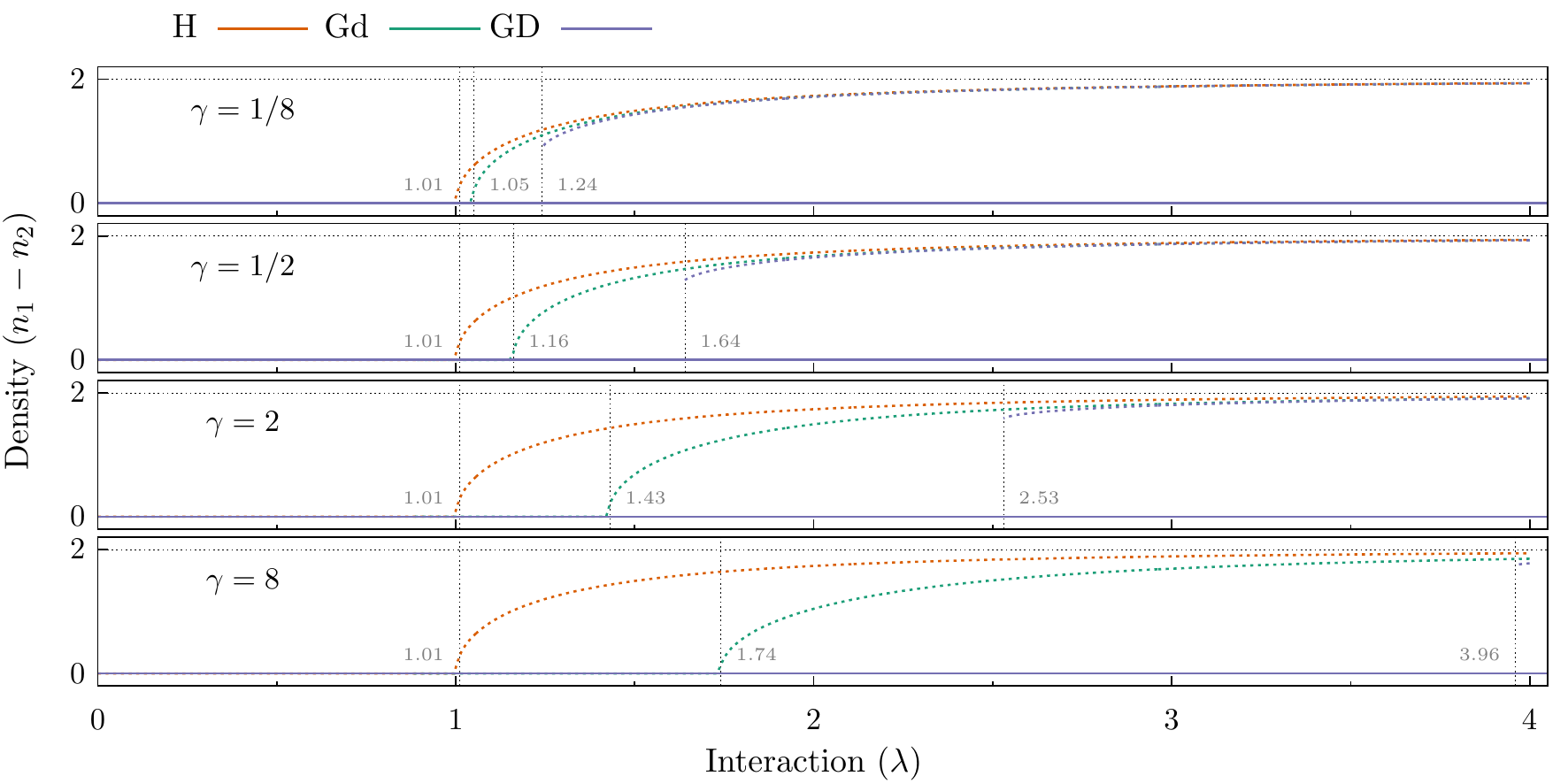}
   \caption{\label{Fig:Densities}
   The relative $n_{1}-n_{2}$ electron density for the symmetric
   (solid lines on top of each other) and asymmetric (dashed lines) solutions
   as a function of the interaction $\lambda$ for different adiabatic
   ratios $\gamma$. Note that the electron density is fixed at two,
   which we find also numerically. The dotted vertical lines denote
   the critical values $\lambda_{C}$ of the interaction given next to a line.
   (color online)}
\end{figure*}

As the first result, we show in Fig.~\ref{Fig:Densities} the relative
electron density $n_{1}-n_{2}$ as a function of the interaction and adiabatic
ratio for the solutions obtained by a symmetric and an asymmetric
initial guess, from here on referred to as asymmetric and symmetric solutions.
The figure shows that a symmetric guess converges to a solution which
has the homogeneous electron density $n_{1}=n_{2}=1$ independent of
the approximation and for all parameters considered. An asymmetric guess
on the other hand converges for a sufficiently strong interaction to a solution
which has an inhomogeneous electron density $n_{1}\neq n_{2}$.
Once asymmetric solutions have emerged as a function of the interaction,
both types of solutions co-exist for the parameters explored in the present work.
The asymmetric solutions have in common that the relative density approaches
two as the interaction gets stronger which, similarly to the mean-field case,
signals the formation of a localized bipolaron. The manner in which this asymmetric solution
appears is however not in common to all approximations.
The mean-field approximation, as discussed earlier, breaks the symmetry
in the manner of a bifurcation. The same is true for the partially
self-consistent approximation, but not for the fully self-consistent
approximation in which asymmetric solutions are found to emerge discontinuously
from the symmetric solution as a function of the interaction.
Additionally, on the contrary to the mean-field case, the value of the interaction
at which the asymmetric solution emerges is not independent of the adiabatic ratio,
but is pushed to higher interactions $\lambda>1$ in the correlated approximations.
\begin{figure}
   \centering
   \includegraphics{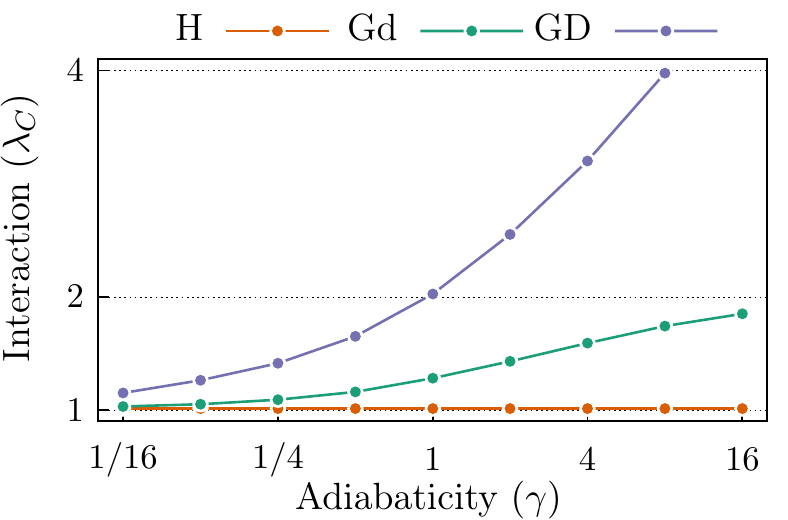}
   \caption{\label{Fig:CriticalPoint}
   The critical value of interaction $\lambda_{C}$ at which an asymmetric
   solution emerges as a function of the adiabatic ratio. (color online)}
\end{figure}
The critical interaction $\lambda_{C}$ below which we do not find
an asymmetric solution is shown in Fig.~\ref{Fig:CriticalPoint}
as a function of the adiabatic ratio. This figure shows that the critical point
approaches one in the adiabatic limit irrespective of the approximation,
while its value depends on the approximation for finite adiabatic ratios
and in the anti-adiabatic limit. As the adiabatic ratio is increased,
the critical interaction approaches two in the partially self-consistent
approximation, which is consistent with the asymptotic limit proposed
in Sec.~\ref{Section:Results:BeyondMeanFieldAsymptoticSolution}.
In the fully self-consistent approximation, the critical point is shifted
to even higher interactions so that for the highest adiabatic ratio
$\gamma=16$ we do not find it in the range of interactions $\lambda\in[0,4]$
studied in this work.

The reason why we find a critical interaction below which we can not obtain
an asymmetric solution is, in the mean-field case, the fact that we do not
allow unphysical complex valued densities. The partially self-consistent
approximation has been shown to reduce to the mean-field approximation
in the adiabatic and anti-adiabatic limits which suggests that
the same reason holds for the partially self-consistent case.
On the other hand, we do not know if this is the case for the fully
self-consistent approximation. The discussion of the phonon vacuum
instability~\cite{alexandrov-1994} given in
Sec.~\ref{Section:Results:PhononVacuumInstability} rather supports the view that 
an asymmetric electron propagator causes a divergence of the phonon propagator,
and due to self-consistency we do not find a physical asymmetric solution.
However, the symmetric solution does not show any signs of the phonon vacuum 
instability in the sense that we find a finite solution.

\begin{figure*}
   \centering
   \includegraphics[width=\textwidth]{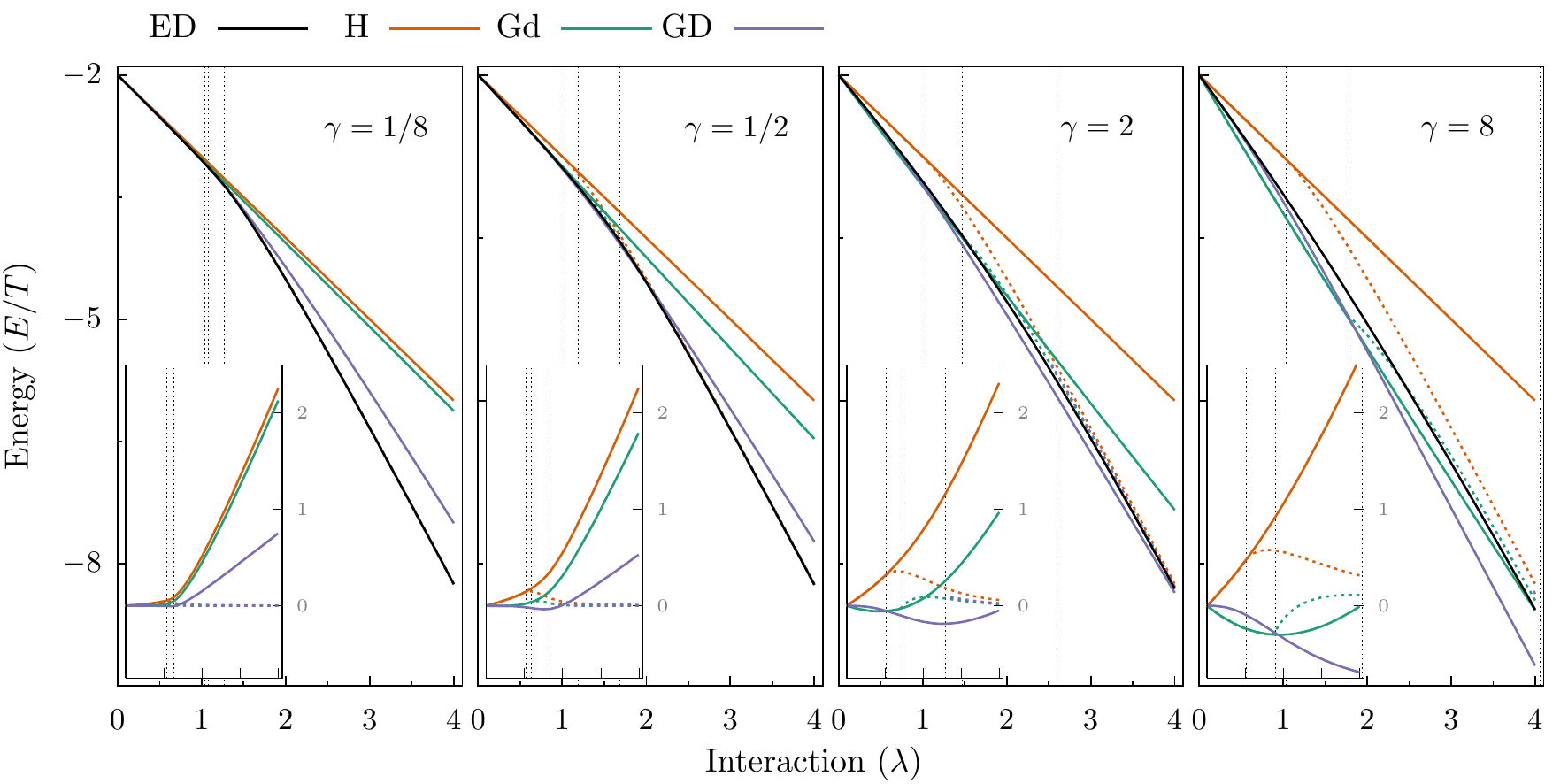}
   \caption{\label{Fig:TotalEnergy}
   The total energies ($E/T$) for the asymmetric (dashed line) 
   and symmetric (solid line) solutions as a function of
   the interaction $\lambda$ for different adiabatic ratios $\gamma$.
   The inset shows the difference $E_{\mathrm{MBPT}}-E_{\mathrm{ED}}$ 
   between approximate (MBPT) and exact (ED) total energies. The dotted vertical
   lines denote the critical values of interaction $\lambda_{C}$.}
\end{figure*}

The total energies shown in Fig.~\ref{Fig:TotalEnergy} in units of the
hopping allow us to determine which one of the solutions,
symmetric or asymmetric, is the lowest energy solution for
a given approximation. These total energies are independent of the adiabatic
ratio in the mean-field approximation, and the asymmetric solution is lower
than the symmetric solution in the energy. The total energies obtained in either
the partially or fully self-consistent approximation are however not independent
of the adiabatic ratio. In the adiabatic limit, we find similarly
to the mean-field situation that the asymmetric solution is lower in energy
once it has emerged. Although this is true for all approximations, adding
self-consistency brings the symmetric solution down in energy,
especially when the phonon propagator is dressed. As the adiabatic ratio
increases, the critical point moves to a higher interaction, and the symmetric
solution is lower in energy for a sufficiently high adiabatic ratio. This
happens when the total energy becomes lower than the exact total energy.
In the anti-adiabatic limit, the symmetric
solution is the lowest energy solution in both correlated
approximations for the interactions considered in this work.

These results confirm that solutions with inhomogeneous densities exist also
in the correlated approximations in which they can be, similarly to
the mean-field case, understood to mimic a bipolaron. The critical interaction
at which these solutions appear agrees qualitatively in the adiabatic limit
with the abrupt formation of a bipolaron in the true ground state, see 
Sec.~\ref{Section:Results:ExactProperties}.
The approximate solutions however break the symmetry suddenly irrespective of
the adiabatic ratio which does not agree with the less abrupt formation of
a bipolaronic ground state in the anti-adiabatic limit. The bottom line is that 
although the asymmetry can be motivated physically, it still does produce 
inhomogeneity in a homogeneous system. This artefact can be circumvented by 
constraining ourselves to the homogeneous solution which we find for all
parameter values considered here.

%------------------------------------------------------------------------------%

\subsection{Comparison With Exact Diagonalization}
\label{Section:Results:ComparisonAgainstExactDiagonalization}
The many-body approximations used here have been shown to exhibit homogeneous
and inhomogeneous densities. The symmetry-broken solution has been attributed
to describe a localized bipolaron, while the nature of the symmetric solution
has so far remained unclear. In the following, we aim to clarify some physical
aspects of both solutions by comparing them energetically with exact
diagonalization.

Let us start by taking a closer look at the exact and approximate 
total energies which are shown in Fig.~\ref{Fig:TotalEnergy}
in units of the hopping. The exact total energies shown
in the main panel display roughly linear behavior as a function of
the effective interaction for the weak and strong interactions.
The slopes of these lines depend on the adiabatic ratio so that for the scaled
energy $E/T$, the slope related to the weak interactions is steeper
in the anti-adiabatic limit. In the adiabatic limit, these two limiting
cases are joined together by an abrupt, smooth change in the slope of
the total energy curve while, in the anti-adiabatic limit, this abrupt change
is smoothened out and moved to higher values of the interaction. This feature
has been shown to correspond to an increase in the binding energy of
the bipolaron, see Sec.~\ref{Section:Results:ExactProperties}, and has been used
to signal a crossover to an on-site bipolaron in the more general
Hubbard-Holstein model~\cite{berciu-2007}.

Let us next compare the approximate total energies to the exact total energy
starting with the symmetric solutions, and using as help the difference
between approximate and exact total energies shown in the inset of
Fig.~\ref{Fig:TotalEnergy}. The symmetric mean-field solution does not
show any change in the slope of the total energy as a function of the interaction,
as seen from Eq.~\eqref{Eq:HartreeTotalEnergy}. This together with the fact
that the total energy is independent of the adiabatic ratio results into
a very poor agreement with the exact total energies for all parameters except
for the weak interaction adiabatic region. The symmetric partially self-consistent
solution shows a change, although almost negligible in comparison 
to the exact case, in its slope as a function of the interaction.
The largest difference to the mean-field case is that the total energy depends
on the adiabatic ratio. This qualitative difference leads to a relatively good 
agreement with the exact results for weak interactions and adiabatic ratios
smaller than one, while in the anti-adiabatic cases the total energy
is underestimated even for a very weak interaction. The fully self-consistent
solution, as opposed to the other symmetric approximate solutions, shows
a clear signature of a change in the slope of the total energy, This change
appears at the correct values of the interaction, but still underestimates
the change of the slope of the exact total energy as a function of $\lambda$.
This observation allows us to conclude that out of the symmetric solutions,
the fully self-consistent solution is in best agreement with the exact results
giving relatively good results for weak- to moderate interactions in
the adiabatic limit and weak interactions in the anti-adiabatic limit.

The asymmetric solutions, unlike their symmetric counterparts, show
all qualitatively similar behavior as a function of the interaction
once they have emerged. The asymmetric solutions appear in the adiabatic regime
roughly at the same point in which the slope of the exact total energy
changes abruptly. As the critical interaction $\lambda_{C}$ is shifted
in the correlated approximations to higher interactions as a function of
the adiabatic ratio, it can be said to follow a similar trend as the slope
of the exact total energy does, although only the symmetry-breaking can be said
to be abrupt in the anti-adiabatic regime.  The total energy differences shown
in the inset illustrate the fact that all symmetry-broken total
energies approach asymptotically to the exact total energy as a function
the interaction. This is particularly true in the adiabatic limit in which
symmetric total energies are in good agreement at the critical values
$\lambda_{C}$, and hence the symmetry-broken total energies are in quantitative
agreement with the exact result. The fact that asymmetric total energies are so
good in the strong interaction limit is related to the almost degenerate ground
and first excited state of the system. A linear combination of these states leads
to a symmetry-broken state whose energy however is extremely close
to the true total energy~\cite{ranninger-1992}.

\begin{figure*}
   \centering
   \includegraphics[width=\textwidth]{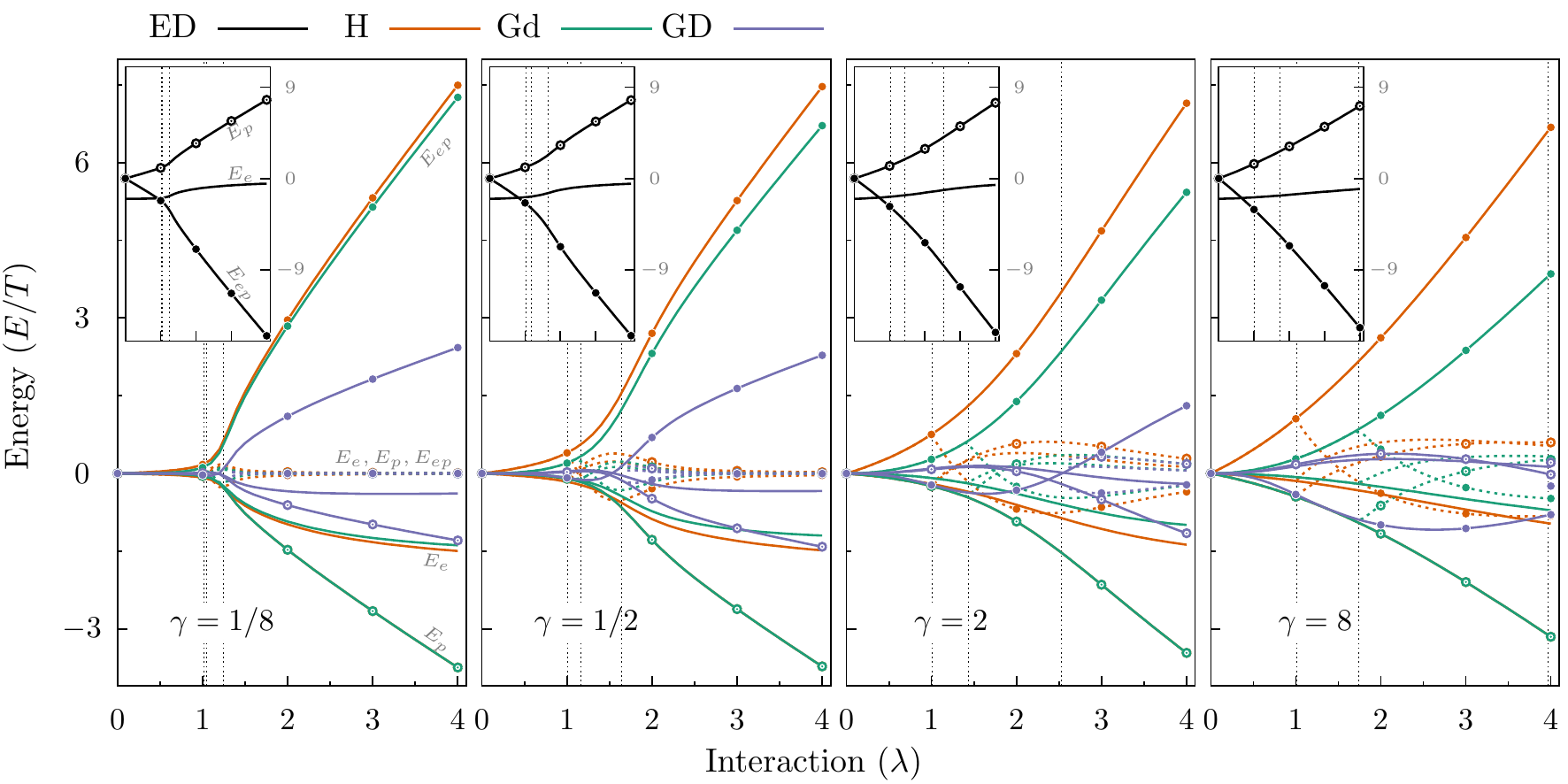}
   \caption{\label{Fig:EnergyComponents}
   The insets display exact phonon (open circles, $E_{p}/T$),
   electron (no points, $E_{e}/T$), and electron-phonon interaction
   (filled circles, $E_{ep}/T$) energies as a function of
   the interaction $\lambda$ for different adiabatic ratios $\gamma$.
   The main panels display the differences $E_{\mathrm{MBPT}}-E_{\mathrm{ED}}$
   between approximate (MBPT) and exact (ED) energy components for
   the symmetric (solid line) and asymmetric (dashed line) solutions.
   The dotted vertical lines denote the critical values of interaction
   $\lambda_{C}$, and labels next to lines on the left most panel are meant
   to guide the reader when moving to the panels on the right. (color online)}
\end{figure*}

The electron, phonon, and electron-phonon interaction energies are shown
for the exact solution in the insets of Fig.~\ref{Fig:EnergyComponents}.
The electron energy is $E_{e}/T=\eta_{2}-\eta_{1}$,
where $\eta_{1}$ and $\eta_{2}$ are eigenvalues of
the spin-summed, one-body reduced density matrix, that is so-called natural
occupation numbers~\cite{lowdin-1956}. This energy contribution remains almost
a constant in the limit of weak interactions, and increases as a function of
the interaction roughly as $-2/\lambda$ in the strong interaction case in
the adiabatic limit. As the adiabatic ratio is increased, the electron energy
changes less abruptly, and shows almost linear increase as a function of
the interaction. The exact phonon energy, which is a sum of the classical
$E_{pC}/T=\lambda$ and quantum $E_{pQ}\equiv E_{p}-E_{pC}$ contributions, and
the electron-phonon interaction energy, which is a sum of the classical
$E_{epC}=-2\lambda$ and quantum $E_{pQ}\equiv E_{ep}-E_{epC}$ contributions,
show a linear increase and decrease as a function of the interaction in the weak
and strong interaction limits, respectively. As for the total energy, the change
from the linear behavior for weak interactions to the linear behavior for strong
interactions appears abruptly in the adiabatic limit, and as smoothened out and
at higher interactions when the adiabatic ratio is increased.

The main panels of Fig.~\ref{Fig:EnergyComponents} display differences between
the approximate and exact energy components as a function of the parameters. Let
us first discuss the symmetric solutions which give the classical energy
components exactly, as seen from Eq.~\eqref{Eq:FieldExpectationValue}, and
therefore the energy differences measure the difference in the quantum
contributions. The overall picture here is that these energy components deviate
from their exact correspondence more than their sum does, that is the total
energy, which implies some kind of error cancellation. The symmetric mean-field
and partially self-consistent solutions describe nuclear motion at
the mean-field level which means that they fail when the quantum contribution
becomes large. Our data shows that the quantum contribution becomes appreciable
for all parameters except for the adiabatic weak interaction regime in which
these two approximations give a relatively good agreement to the exact phonon
energy. In the fully self-consistent case, we however dress the phonon
propagator, and therefore gain the ability to capture some of the quantum
contribution. This is seen in the figure as an improved agreement with the exact
phonon energy for all adiabatic ratios and values of the interactions.
The symmetric mean-field solution gives $-2T$ as the electron energy which means
that it agrees adequately with exact results only in the adiabatic weak
interaction limit in which the exact electron energy remains roughly constant as
a function of the interaction. The partially self-consistent approximation
incorporates some electron correlation beyond mean-field which is seen as
improved agreement with exact results for all adiabatic ratios. The qualitative
gain of further introducing self-consistency in the fully self-consistent case
is that the electron energy is in a better agreement with the exact result also
in the adiabatic limit. The last energy component is the electron-phonon
interaction energy which deviates the most from the exact result. As in the case
of other energy components, the mean-field solution again fails to give a good
agreement with exact results except in the adiabatic weak coupling regime.
The partially and especially the fully self-consistent solutions improve on
the mean-field picture, so that the latter is again in the best agreement with
exact electron-phonon interaction energy for all parameters considered.
The error in the interaction energy is of the opposite sign to the error in
the phonon energy, mostly also to the electron energy, and therefore leads to
the error cancellation observed for the total energy.

\begin{figure}
   \centering
   \includegraphics{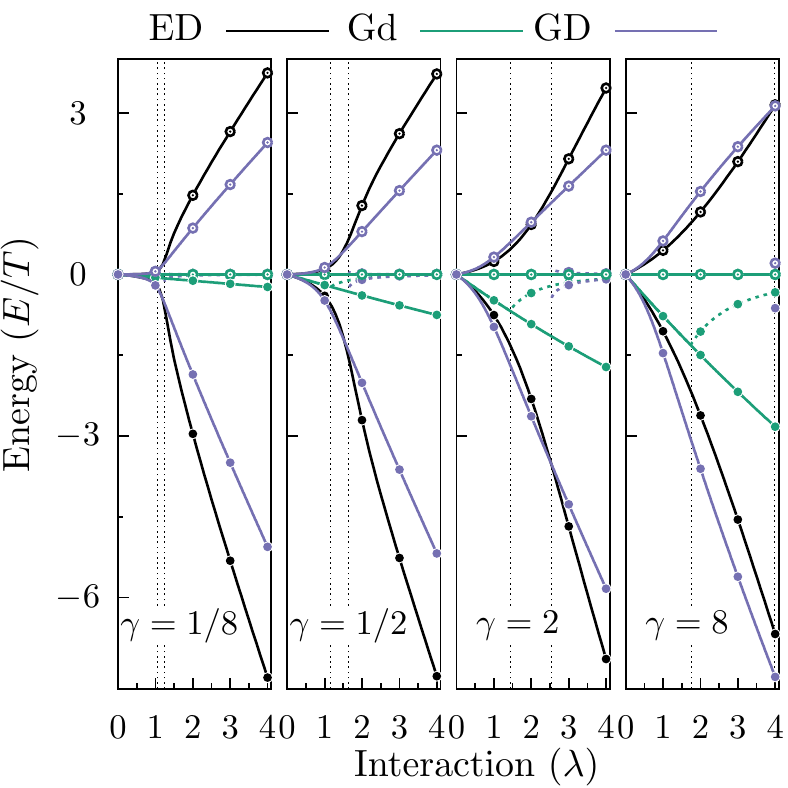}
   \caption{\label{Fig:AsymmetricEnergyComponents}
   The quantum or correlation contributions to the phonon (open circles,
   $E_{pQ}/T$) and electron-phonon interaction (filled circles, $E_{epQ}/T$)
   energies for the exact, and asymmetric (dashed line) and symmetric
   (solid line) solutions as a function of the interaction $\lambda$ for
   different adiabatic ratios $\gamma$. The dotted vertical lines denote
   the critical values of interaction $\lambda_{C}$. (color online)}
\end{figure}

Once the asymmetric solutions appear when the interaction is increased, their
energy components behave similarly to one another, as seen for the total
energies. Also the trend is the same as for the total energies, that is all
energy components approach asymptotically as a function of interaction towards
the corresponding exact energy component. There is even a quantitative
agreement with the exact energy components for a high enough interaction.
Additionally, we find, like in the symmetric case, that there is an error
cancellation occurring between the different components. The quantum
contributions to the phonon and electron-phonon interaction energies are shown
in Fig.~\ref{Fig:AsymmetricEnergyComponents} for the correlated solutions,
and these contributions are by construction identically to zero for the mean-field
approximation. This figure shows that an asymmetric solution, irrespective of
the approximation, gives quantum energy contributions which tend towards zero as
a function of the interaction, and hence the energies consist solely of
a classical-like, mean-field contribution.

\begin{figure}
   \centering
   \includegraphics{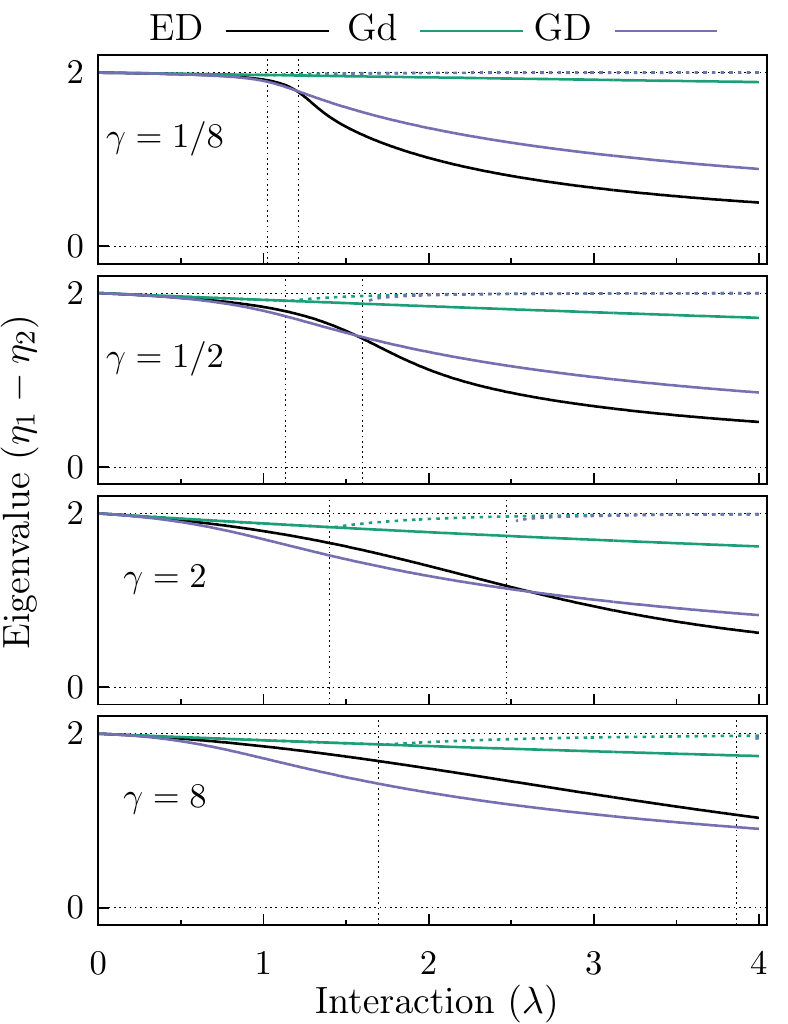}
   \caption{\label{Fig:NaturalOccupationNumbers}
   The difference $\eta_{1}-\eta_{2}$ of
   the eigenvalues of the spin-summed, electron one-body reduced density matrix
   for the asymmetric (dashed line)
   and symmetric (full line) solutions as a function of
   the interaction ($\lambda$) for different adiabatic ratios
   ($\gamma$). The dotted vertical lines denote
   the critical values of interaction $\lambda_{C}$. (color online)}
\end{figure}

As a last indicator of the physical nature these solutions, we show in
Fig.~\ref{Fig:NaturalOccupationNumbers} the difference $\eta_{1}-\eta_{2}$ of
the eigenvalues of the spin-summed, electron one-body reduced density matrix.
This difference is only shown for the exact solution and the correlated
approximations since the natural occupation numbers are by construction fixed to
$\eta_{1}=2$ and $\eta_{2}=0$ in the mean-field
approximation. The figure shows that the difference approaches two as a function
of the interaction also for the asymmetric solutions of the correlated
approximations. This means that the agreement with the exact electron energy
follows through the symmetry-broken eigenvectors of the density matrix. However,
this is not the case for the symmetric solutions out of which the fully
self-consistent agrees qualitatively with the exact natural occupation numbers.
There is also an observation to be made from this figure, namely due to the fact
that the particle number is always two, the eigenvalues are between zero and two
for all approximate solutions. This shows that the many-body approximation here
satisfy an $N$-representability condition~\cite{davidson-book}.
In the anti-adiabatic limit, the natural occupation numbers can be used to
construct the ground-state
wave function, and therefore they give invaluable information about
the bipolaronic ground state. The first step to make this correspondence is to
map  the Holstein model to the attractive Hubbard model using the Lang-Firsov
transformation~\cite{stefanucci-book,macridin-2004}. The Hamiltonian is then
given for two electrons by
\begin{align*}
   \frac{\hat{H}}{T}
   =& -\lambda
   +\gamma\sum_{i=1}^{2}\hat{\tilde{a}}_{i}^{\dagger}\hat{\tilde{a}}_{i}
   -\lambda\sum_{i=1}^{2}\hat{\tilde{n}}_{i\uparrow}
   \hat{\tilde{n}}_{i\downarrow} \notag\\
   &-\sum_{\sigma}\big(
   \hat{X}_{1}^{\dagger}\hat{X}_{2}
   \hat{\tilde{c}}_{1\sigma}^{\dagger}\hat{\tilde{c}}_{2\sigma}
   +\hat{X}_{2}^{\dagger}\hat{X}_{1}
   \hat{\tilde{c}}_{2\sigma}^{\dagger}\hat{\tilde{c}}_{1\sigma}\big)\, ,
\end{align*}
where the transformed operators are
\begin{align*}
   \hat{\tilde{a}}_{i}
   &\equiv \hat{a}_{i}+\sqrt{\lambda/2\gamma}\hat{n}_{i}\, ,\\
   \hat{\tilde{c}}_{i}
   &\equiv \hat{X}_{i}\hat{c}_{i}\, ,\\
   \hat{X}_{i}
   &\equiv e^{\sqrt{\lambda/2\gamma}(\hat{a}_{i}^{\dagger}-\hat{a}_{i})}\, ,
\end{align*}
and the density operator
$\hat{\tilde{n}}_{i\sigma}\equiv\hat{\tilde{c}}^{\dagger}_{i\sigma}\hat{\tilde{c}}_{i\sigma}$
is defined as usual. If we take the anti-adiabatic limit, or more precisely
$\lambda/2\gamma\rightarrow 0$ such that $\lambda$ remains finite~\cite{macridin-2004},
we arrive at the Hamiltonian
\begin{align*}
   \hat{H}_{\mathrm{H}}
   \equiv& -T\lambda
   +\omega_{0}\sum_{i=1}^{2}\hat{a}_{i}^{\dagger}\hat{a}_{i}
   -\lambda\sum_{i}\hat{n}_{i\uparrow}\hat{n}_{i\downarrow} \notag\\
   -&T\sum_{\sigma}\big(\hat{c}_{2\sigma}^{\dagger}\hat{c}_{1\sigma}
   +\hat{c}_{1\sigma}^{\dagger}\hat{c}_{2\sigma}\big)\, ,
\end{align*}
where the phonon part is just the non-interacting one, and the electronic
part is equal to the Hamiltonian operator of the attractive Hubbard model
with the interaction $-\lambda$. The ground-state wave function is a product
of a zero-phonon state and an electronic wave function which can be written
by L\"{o}wdin-Shull expansion~\cite{lowdin-1956} as
\begin{align*}
   \lvert \Psi_{e}\rangle
   =&\frac{b_{1}+b_{2}}{2}
   \big(\hat{c}^{\dagger}_{1\uparrow}\hat{c}^{\dagger}_{1\downarrow}
   +\hat{c}^{\dagger}_{2\uparrow}\hat{c}^{\dagger}_{2\downarrow}\big)
   \lvert 0\rangle_{e}\notag\\ 
   &+\frac{b_{1}-b_{2}}{2}
   \big(\hat{c}^{\dagger}_{1\uparrow}\hat{c}^{\dagger}_{2\downarrow}
   +\hat{c}^{\dagger}_{2\uparrow}\hat{c}^{\dagger}_{1\downarrow}\big)
   \lvert 0\rangle_{e} \, ,
\end{align*}
where $\lvert 0\rangle_{e}$ is the empty electronic state, and
$b_{i}=\sqrt{\eta_{i}/2}$ are called natural amplitudes. In general
the coefficients $b_i$ are defined up to a sign~\cite{giesbertz-2013a,
giesbertz-2013b}. However, since in this case the ground state can be calculated
analytically we can readily determine the signs to be positive in our case.
At the limit of strong interactions, the natural occupation numbers approach one
which implies that $b_{1}-b_{2}$ approaches zero and therefore the ground-state
wave function describes two electrons on the same site, which is the bipolaronic
ground state in the anti-adiabatic case. The natural occupation numbers shown
for the approximations in the anti-adiabatic case $\gamma=8$ then imply
that the fully self-consistent approximation describes a paired two-electron
state without symmetry-breaking and localization. The figure also shows that
this pairing is overestimated for a wide range of values of the interaction
in the intermediate to strong couplings.

The presented results re-enforce the picture that the asymmetric solutions can
be seen to describe a classical-like, localized bipolaron. This kind of state is
the degenerate ground-state solution in the extreme adiabatic and anti-adiabatic
limits obtained by neglecting the nuclear and electron kinetic energies, but not
for finite parameters for which degeneracy is broken and the solution is
symmetric. On the other hand, we have seen that if we restrict ourselves to
a solution which respects this symmetry, only the fully self-consistent
approximation shows sufficient indirect evidence that a correlated two-electron
state is formed as a function of the interaction. This indirect evidence is
further supported by the natural occupation number analysis in
the anti-adiabatic limit.

%%%%%%%%%%%%%%%%%%%%%%%%%%%%%%%%%%%%%%%%%%%%%%%%%%%%%%%%%%%%%%%%%%
%                         Conclusion                             %
%%%%%%%%%%%%%%%%%%%%%%%%%%%%%%%%%%%%%%%%%%%%%%%%%%%%%%%%%%%%%%%%%%

\section{Conclusion}
\label{section:Conclusion}
We have studied the ground-state properties of the homogeneous two-site and
two-electron Holstein model. Our study has been conducted using many-body
perturbation theory, based on the Hartree, and partially and fully
self-consistent Born approximations. We have calculated electron densities,
natural occupation numbers, as well as total energies and electron, phonon and
electron-phonon interaction energies. We have analyzed the results by analytic
and numerical means, and compared them to numerically exact results obtained
via exact diagonalization. 

The results show that there exists a critical interaction above which
the many-body approximations support at least three solutions. One of these
solutions is spatially homogeneous while the other two are inhomogeneous in both
electron population and nuclear displacement, and therefore break the reflection
symmetry of the system. The symmetry-broken electron density approaches two on
one of the sites and zero on the other, while the displacement increases and
decreases linearly with the electron density, respectively. The energy
components and total energies of these solutions have been shown to agree at
best quantitatively with exact results, while their quantum contributions
approach zero for strong interactions. These observations support the physical
picture that the inhomogeneous solutions represent a localized, classical-like
bipolaron. The asymmetric solutions are alike irrespective of the approximation,
but this is not to the case for the homogeneous solutions. The comparison of
the energy components and total energies shows that the homogeneous solutions of
the Hartree and partially self-consistent Born approximations do not agree even
qualitatively with exact results for intermediate to strong interactions. The
fully self-consistent approximation instead does compare well with exact results
at least up to intermediate interactions in the adiabatic case, and at least for
weak interactions in the anti-adiabatic case. This approximation also shows a 
significant, although underestimated, change in the slope of the total energy in
the adiabatic case $\gamma=1/8$ in a qualitative agreement with the exact
solution where we associate this change with a crossover to a bipolaronic state
as in~\cite{berciu-2007}. The homogeneous solution of the fully self-consistent
Born approximation is then understood to describe partially a bipolaron
crossover in this adiabatic case. A study in an infinite dimensional extended
system has shown that this approximation predicts qualitatively a polaronic
crossover, but gives only exponentially decaying quasi-particle spectral weight
and hence no true bipolaronic metal-insulator transition~\cite{capone-2003}.
The spin-summed natural occupation  numbers obey a trend analogous to the
energetics, they approach each other as a function of the interaction in the
exact and fully self-consistent Born solutions but do not change as
significantly in the less sophisticated approximations. We have shown by mapping
the model to the attractive Hubbard model in the anti-adiabatic limit that
natural occupation numbers which approach one another indicate the formation of
a paired two-electron state. This observation favors a conclusion that the
homogeneous solution of the fully self-consistent Born approximation describes
a paired two-electron state for the anti-adiabatic case $\gamma=8$ considered
here.

The presented results contribute to understanding how these approximations
behave in the presence of a strong localizing interaction. In particular,
we have shown that the many-body approximations used in this work have
multiple physical solutions, and they give rise to spontaneous symmetry-breaking
if allowed. These general features appear naturally in the bipolaronic system
studied here, and therefore raise a question if these features manifest
themselves in dynamical settings, such as in time-dependent quantum transport
through molecular junctions, as instable or metastable dynamics. The extension
of the present formalism to these cases requires the solution of the two-time
Kadanoff-Baym equations for open systems which we have studied extensively in
the last few years for the case of purely electronic systems. The next step into
this direction is to study how dynamical properties at the level of response
functions are described in the many-body approximations discussed
here~\cite{saekkinen-2014b}. This knowledge will be of great importance for understanding how the new time-scale related to nuclear motion appears in the 
transient regime~\cite{perfetto-2013}. Lastly, we remark that this may also
require further simplifications of the Kadanoff-Baym equations which lead to
computationally more favorable single-time equations~\cite{pavlyukh-2009,
pavlyukh-2011,marini-2013,balzer-2013,latini-2014}.

%------------------------------------------------------------------------------%

\begin{acknowledgements}
We acknowledge CSC – IT Center for Science Ltd. and the Fritz-Haber-Institute 
for the allocation of computational resources. RvL acknowledges the Academy of 
Finland for support under grant No.~267839.
\end{acknowledgements}

%------------------------------------------------------------------------------%
\newpage

\appendix

%%%%%%%%%%%%%%%%%%%%%%%%%%%%%%%%%%%%%%%%%%%%%%%%%%%%%%%%%%%%%%%%%%
%                       Numerics                                 %
%%%%%%%%%%%%%%%%%%%%%%%%%%%%%%%%%%%%%%%%%%%%%%%%%%%%%%%%%%%%%%%%%%

\section{Energy Functional}
\label{Appendix:EnergyFunctional}
The total energies presented in this work are obtained using an energy
functional whose derivation will be given below. The underlying idea is that the
electron propagator obeys the equation of motion
\begin{align*}
   \imath\partial_{z}G_{ij}(z;z')
   &=\delta_{ij}\delta(z,z')\notag\\
   &+\sum_{k}h_{ik}G_{kj}(z;z')\notag\\
   &-\imath\sum_{kl}M_{ik}^{l}
   \expval{\toop\left\{\hat{\phi}_{l}(z)
   \hat{c}_{H;k}(z)\hat{c}^{\dagger}_{H;j}(z')\right\}}
   \, ,
\end{align*}
which together with the phonon propagator allow us to
write the energy components as
\begin{align*}
   E_{p}(z)
   &= \sum_{IJ}\Omega_{IJ}\expval{\hat{\phi}_{H;I}(z)\hat{\phi}_{H;J}(z)}
   \notag\\
   &= -\frac{1}{\imath}\sum_{IJ}\Omega_{IJ}\big(D_{JI}(z,z^{+})
   -\imath\phi_{I}(z)\phi_{J}(z)\big)
   \, ,\\
   E_{e}(z)
   &= \sum_{ij}h_{ij}\expval{\hat{c}^{\dagger}_{H;i}(z)\hat{c}_{H;j}(z)}
   \notag\\
   &=\frac{1}{\imath}\sum_{ij}h_{ij}G_{ji}(z,z^{+})
   \, ,\\
   E_{ep}(z)
   &= \sum_{I}\sum_{jk}M_{jk}^{I}
   \expval{\hat{\phi}_{H;I}(z)\hat{c}_{H;j}^{\dagger}(z)\hat{c}_{H;k}(z)}
   \notag\\
   &= \frac{1}{\imath}\sum_{ik}\big(\imath\delta_{ik}\partial_{z}
   -h_{ik}\big)G_{ki}(z;z')\big|_{z'=z^{+}}
   \, ,
\end{align*}
where the subscript $H$ denotes a Heisenberg picture operator. Finally, by
combining these results, we find that the total energy can be written as
\begin{align*}
   E_{\mathrm{GM}}(z)
   &= \sum_{i}\partial_{z}G_{ii}(z;z')\big|_{z'=z^{+}} \notag\\
   &-\frac{1}{\imath}\sum_{IJ}\Big(\Omega_{IJ}\big(D_{JI}(z,z^{+})
   -\imath\phi_{I}(z)\phi_{J}(z)\big)\Big) \, ,
\end{align*}
and is therefore a functional of the electron and phonon propagators. We call
this functional according to its deriving principles Galitski-Migdal
functional~\cite{galitskii-1958,almbladh-1999,holm-2004}.

%------------------------------------------------------------------------------%

\section{Numerical Details}
\label{Appendix:NumericalDetails}
The numerical method used in the present work to solve the coupled,
equilibrium Dyson equations is the widely used method of fixed-point iterations.
This method has been previosly described in~\cite{Stan-2009}.
Here we extend this approach with some necessary modifications to allow
for electron-phonon coupling in the system.

In the present work, the electron and phonon propagators are discretized
in the imaginary-time $\tau$ on the geometric grid 
\begin{align*} 
   \tau_{k+1}-\tau_{k}
   \equiv a b^{k}
\end{align*}
such that $\tau_{0}=0$ and $\tau_{k}\in[0,\beta/2]$.
The grid in the interval $\tau\in[\beta/2,\beta]$ is obtained by mirroring
of this grid with respect to $\beta/2$. The grid parameters used here are
$a=10^{-3}$ and $b=1.1$, or even smaller.
The equations are first rewritten in terms of static
electron $g_{\mathrm{s}}^{M}$ and phonon $d_{\mathrm{s}}^{M}$ propagators
which satisfy the equations of motion
\begin{align*}
   -\big(\m{\partial}_{\tau}+\m{h}-\mu_{\mathrm{s}}\big)
   \m{g}_{\mathrm{s}}^{M}(\tau)
   &=\imath\m{\delta}(\tau)
   +\big[\m{\Sigma}_{\mathrm{s}}^{M}
   \star\m{g}_{\mathrm{s}}^{M}\big](\tau)
   \, ,\\
   -\big(\m{\alpha}\partial_{\tau}+\m{\omega}\m{1}\big)
   \m{d}_{\mathrm{s}}^{M}(\tau)
   &=\imath\m{\delta}(\tau)
   +\big[\m{\Pi}_{\mathrm{s}}^{M}
   \star\m{d}_{\mathrm{s}}^{M}\big](\tau) \, ,
\end{align*}
where $\mu_{\mathrm{s}}$ denotes a static chemical potential, 
while $\Sigma_{\mathrm{s}}^{M}$ and $\Pi_{\mathrm{s}}^{M}$ denote static
self-energies. The word static refers here to the fact that these quantities
are independent of the propagators $G^{M}$ and $D^{M}$ which we want to solve.
The Dyson equations for these propagators can be then rewritten as
\begin{align*}
   \m{G}^{M}(\tau)
   &=\m{g}_{\mathrm{s}}^{M}(\tau)
   +\big[\m{g}_{\mathrm{s}}^{M}
   \star\m{\overline{\Sigma}}^{M}
   \star\m{G}^{M}\big](\tau)
   \, ,\\
   \m{D}^{M}(\tau)
   &=\m{d}_{\mathrm{s}}^{M}(\tau)
   +\big[\m{d}_{\mathrm{s}}^{M}
   \star\m{\overline{\Pi}}^{M}
   \star\m{D}^{M}\big](\tau) \, ,
\end{align*}
where we defined the effective electron
$\m{\overline{\Sigma}}_{S}^{M}(\tau)
\equiv\m{\Sigma}^{M}(\tau)-\m{\Sigma}_{\mathrm{s}}^{M}(\tau)
+\m{\delta}(\tau)\big(\mu-\mu_{\mathrm{s}}\big)$
and phonon
$\m{\overline{\Pi}}_{S}^{M}(\tau)
\equiv\m{\Pi}^{M}(\tau)-\m{\Pi}_{\mathrm{s}}^{M}(\tau)$ self-energies.

The method of fixed-point iterations relies on the idea that the solution
to the equations is a fixed-point of a mapping which given 
the propagators as an input gives new propagators as its output.
This procedure can be iterated to generate a sequence of propagators
which in the ideal case converges to a fixed point of this mapping.
This work is based on a scheme in which the propagators
of the $k$-th iteration are given as inputs to the mappings
\begin{align*}
   \mathcal{F}_{G}:\;
   \{G,D\}
   &\rightarrow
   \m{g}_{\mathrm{s}}^{M}
   +\m{g}_{\mathrm{s}}^{M}
   \star\m{\overline{\Sigma}}[G,D]
   \star\m{G}
   \, ,\\
   \mathcal{F}_{D}:\;
   \{G,D\}
   &\rightarrow
   \m{d}_{\mathrm{s}}^{M}(\tau)
   +\m{d}_{\mathrm{s}}^{M}
   \star\m{\overline{\Pi}}[G]
   \star\m{D} \, ,
\end{align*}
which give a new pair of propagators as its output. This mapping by itself
can be unstable, so in order to improve the convergence of this sequence,
we have implemented a well-known convergence acceleration method called
Direct Inversion of Iterative Subspace (DIIS)~\cite{pulay-1980,thygesen-2008}.
We apply this technique by storing a history
of $N$ latest iterates, and constructing new optimized electron
and phonon propagators in the $k$-th iteration round according to
\begin{align*}
   \m{G}_{\mathrm{opt}}^{k}(\tau)
   &=\sum_{\substack{i=0\\i\leq k}}^{N}c_{i}\m{G}_{k-i}(\tau) \, ,\\
   \m{D}_{\mathrm{opt}}^{k}(\tau)
   &=\sum_{\substack{i=0\\i\leq k}}^{N}c_{i}\m{D}_{k-i}(\tau) \, ,
\end{align*} 
where the coefficients $c_{i}$ are fixed by minimizing the residual norm
$r_{k}\equiv\norm{\m{G}_{\mathrm{opt}}^{k}-\mathcal{F}_{G}\big[\m{G}_{\mathrm{opt}}^{k}\big]}^{2}
+\norm{\m{D}_{\mathrm{opt}}^{k}-\mathcal{F}_{D}\big[\m{D}_{\mathrm{opt}}^{k}\big]}^{2}$
under the constraint that the coefficients $c_{i}$ sum up to unity.
The norm is defined in the usual way $\norm{\m{a}}\equiv\sqrt{\langle \m{a},\m{a}\rangle}$
in terms of the inner product
\begin{align*}
   \langle \m{a},\m{b}\rangle
   &\equiv\int\limits_{0}^{\beta}\!d\tau\;
   \tr\big(\m{a}^{\dagger}(\tau)\m{b}(\tau)\big)
\end{align*}
where $\m{a},\m{b}$ are matrix valued Matsubara functions, and $\tr$ denotes the usual
matrix trace. To overcome the fact that this optimization problem is inherently
non-linear, we make the usual approximation by assuming that the mappings
are approximately linear which allows us to write
\begin{align*}
   r_{k}
   &\approx\sum_{\substack{i,j=0\\i,j\leq k}}^{N}c_{i}^{\dagger}c_{j}
   r_{k,i}r_{k,j}
\end{align*}
where we defined the residual norm
$r_{k,i}\equiv\norm{\m{G}_{k-i}-\mathcal{F}_{G}\big[\m{G}_{k-i}\big]}^{2}
+\norm{\m{D}_{k-i}-\mathcal{F}_{D}\big[\m{D}_{k-i}\big]}^{2}$.
This simplified constrained optimization problem leads
to the usual set of linear equations
\begin{align*}
   \begin{pmatrix}
      R_{k}^{*}R_{k}^{T} & -1 \\
      -1 & 0
   \end{pmatrix}
   \begin{pmatrix}
      C \\
      \mu
   \end{pmatrix}
   =
   \begin{pmatrix}
       \bar{0} \\
      -1
   \end{pmatrix}
\end{align*}
where $C\equiv (c_{0},\dots,c_{N})^{T}$ and
$R_{k}\equiv(r_{k,0},\dots,r_{k,N})^{T}$ are vectors, $T$ denotes the vector
transpose, and $\mu$ is a Lagrange multiplier enforcing the equality constraint.
Finally, new iterates are given by the linear combinations
\begin{align*}
   \m{G}_{k+1}^{M}
   &=d\mathcal{F}_{G}[\m{G}_{k}^{M}]
   +(1-d)\m{G}_{\mathrm{opt}}^{k}\, ,\\
   \m{D}_{k+1}^{M}
   &=d\mathcal{F}_{D}[\m{D}_{k}^{M}]
   +(1-d)\m{D}_{\mathrm{opt}}^{k} \, ,
\end{align*}
where $d$ is an empirically chosen damping factor,
which is typically fixed to $0.2-1.0$.
Although convergence of this iterative sequence is not guaranteed,
we have found that it does usually converge to a solution. The convergence
criteria used in the present work is that the change of the total energy 
and the value of the residual norm $r_{k,0}$ are both below
the tolerance $10^{-8}$. 

%%%%%%%%%%%%%%%%%%%%%%%%%%%%%%%%%%%%%%%%%%%%%%%%%%%%%%%%%%%%%%%%%%
%                        Bibliography                            %
%%%%%%%%%%%%%%%%%%%%%%%%%%%%%%%%%%%%%%%%%%%%%%%%%%%%%%%%%%%%%%%%%%


\end{document}